\documentclass[epsfig,11pt,amsmath,tightenlines,onecolumn]{revtex4}
\usepackage{graphicx}

\begin{document}

\title{Reaction Brownian Dynamics and the effect of spatial fluctuations on the
gain of a push-pull network}

\author{Marco J. Morelli}
\affiliation{FOM Institute for Atomic and Molecular Physics, Kruislaan 407, 1098 SJ Amsterdam, The Netherlands}

\author{Pieter Rein ten Wolde}
\email{tenwolde@amolf.nl}
\affiliation{FOM Institute for Atomic and Molecular Physics, Kruislaan 407, 1098 SJ Amsterdam, The Netherlands}

\date{\today}

\begin{abstract}

  Brownian Dynamics algorithms are widely used for simulating
  soft-matter and biochemical systems.  In recent times, their
  application has been extended to the simulation of coarse-grained
  models of cellular networks in simple organisms. In these models,
  components move by diffusion, and can react with one another upon
  contact. However, when reactions are incorporated into a Brownian
  Dynamics algorithm, attention must be paid to avoid violations of
  the detailed-balance rule, and therefore introducing systematic
  errors in the simulation. We present a Brownian Dynamics algorithm
  for reaction-diffusion systems that rigorously obeys detailed
  balance for equilibrium reactions. By comparing the simulation
  results to exact analytical results for a bimolecular reaction, we
  show that the algorithm correctly reproduces both equilibrium and
  dynamical quantities.  We apply our scheme to a ``push-pull''
  network in which two antagonistic enzymes covalently modify a
  substrate.  Our results highlight that the diffusive behaviour of
  the reacting species can reduce the gain of the response curve of
  this network.

\end{abstract}

\maketitle

\section{Introduction}
Most, if not all, biological processes are regulated by biomolecules,
such as proteins and DNA, which chemically and physically interact
with one another in what are called biochemical networks. These
networks are often highly non-linear, which means that mathematical
modelling is critical for understanding and predicting their
behaviour. The dominant paradigm has been to consider the living cell
to be a spatially homogeneous environment, analogous to a well-stirred
reactor. It is increasingly recognised, however,
that the cell is a highly inhomogeneous environment, in which
compartmentalisation, scaffolding and localised interactions are
actively exploited to enhance the regulatory function of biochemical
networks. This means that it becomes important to not only describe
the biochemical network in time, but also in space.

In this manuscript, we present an algorithm for simulating biochemical
networks in time and space that is based upon Brownian
Dynamics. Brownian Dynamics is a stochastic dynamics scheme, in which
the solvent is treated implicitly; only solutes are described
explicitly. The forces experienced by the solutes contain a
contribution from the interactions with the other solutes and a random
part, which is the dynamical remnant of the collisions with the
solvent molecules.  A pioneering BD algorithm was introduced by Ermak
and McCammon \cite{Ermak78}, and detailed, atomistic Brownian Dynamics
simulations have been performed to study the dynamics of
enzyme-substrate and protein-protein association reactions
\cite{Northrup92,Wade96,Huber96,Gabdoulline97,Zou00,Gabdoulline01,Gabdoulline02}. More
recently, Brownian Dynamics has not only been used to study the
association between two proteins, but also to simulate networks of
interacting biomolecules \cite{VirtualCell,Smoldyn,Mcell}. In order to
simulate these large systems at the biologically relevant length and
time scales, molecules are coarse-grained to the level of simple
geometrical objects, which can diffuse and react with other chemical
species in a confined geometry.

While Brownian Dynamics algorithms for simulating biochemical networks
are based upon a simplified description of the molecules and their
interactions, they do go beyond the conventional kinetic Monte Carlo
schemes to simulate biochemical networks \cite{Gillespie77}. These
algorithms are based upon the zero-dimensional chemical master
equation, and, as such, they take into account the discrete nature
of the components and the stochastic character of their interactions. However,
they assume that at each instant in time the particles are uniformly
distributed in space.  In contrast, Brownian Dynamics based algorithms
take into account not only the particulate nature of the molecules and
the probabilistic character of their interactions, but also that at
any moment in time the particles may be non-uniformly distributed in
space. Brownian Dynamics thus accounts for both temporal and spatial
fluctuations of the components. Moreover, it allows for spatial
gradients and for localised interactions in the network. 

Recently, a number of stochastic techniques have been developed that
make it possible to simulate biochemical networks in time and
space. Some techniques are based upon Brownian Dynamics
\cite{VirtualCell,Smoldyn,Mcell}, while others are based upon the
reaction-diffusion master equation
\cite{MesoRD,SmartCell,Lemerle05,Rodriguez06}. The advantage of
Brownian Dynamics based techniques is that they are truly
particle-based, which means that they do not have to rely on a
mesoscopic length and time scale on which the system is
well-stirred. We have recently developed an entirely novel algorithm,
called Green's Function Reaction Dynamics (GFRD), which is
particle-based scheme to simulate biological networks in time and
space, like Brownian Dynamics. However, in contrast to Brownian
Dynamics, GFRD is an event-driven algorithm, which uses Green's
functions to concatenate the propagation of the particles in space
with the chemical reactions between them. This makes GFRD orders of
magnitude more efficient than Brownian Dynamics when the
concentrations are below $0.1\,-\,1\mu{\rm M}$. For higher
concentrations, or for reactions near surfaces, brute-force Brownian
Dynamics is more efficient, because of the smaller computational
overhead per time step.

Although the main idea of applying Brownian Dynamics to
reaction-diffusion systems is straightforward, a number of ingredients
has to be examined with care. One is: to which processes do the
association and dissociation rates as used in the simulations
correspond to? To the intrinsic rates, which are the reaction rates at
contact, or to the effective rates that also take into account the effect
of diffusion? The other issue is detailed balance. Biochemical
networks often contain reactions that do not consume energy or are
otherwise driven out of equilibrium. These equilibrium reactions
should obey detailed balance. Even though a number of Brownian
Dynamics based algorithms have been presented
\cite{VirtualCell,Smoldyn,Mcell}, this question has, to our knowledge,
not been systematically addressed.

In this paper, we present a Brownian Dynamics algorithm that
rigorously obeys detailed balance and is thus able to reproduce the
equilibrium properties of a reaction-diffusion system. In
Section~\ref{sec:methods_BD}, we derive our algorithm on the basis of
the statistical mechanics of chemical reactions. The algorithm is
subjected to stringent tests in Section~\ref{sec:tests}: besides
equilibrium properties, we test also how well the algorithm reproduces
the dynamical behavior of a bimolecular reaction, for different values
of the time step $\Delta t$.  A comparison with a stochastic algorithm
that does not account for spatial fluctuations is also
presented. Finally, in Section~\ref{sec:p-p} we show an illustrative
application of our algorithm to a simple coarse-grained model of a
chemical species under the action of two enzymes operating in opposite
directions (the so-called ``push-pull'' model system). Simulations
conducted with our BD algorithm show that both spatial and temporal
fluctuations reduce the gain of the response of the system.

\section{Methods}\label{sec:methods_BD}

\subsection{Detailed balance}
Before we present the outline of the algorithm in the next section, we
discuss the detailed-balance rule that must be obeyed for
equilibrium reactions. To this end, we will consider the elementary
bimolecular reaction:
\begin{equation}\label{eq:A+B=C}
A+B \rightleftharpoons C\qquad (k_{\rm on},k_{\rm off}).
\end{equation}
Here $k_{\rm on}$ is the macroscopic forward rate for the association
of molecules $A$ and $B$, and $k_{\rm off}$ is the macroscopic backward rate for
their dissociation. 
The macroscopic expression for the equilibrium constant for this reaction is
\begin{equation}
K_{\rm eq}=\frac{k_{\rm on}}{k_{\rm off}}=\frac{[C]}{[A][B]},
\end{equation}
where $[X]$ is the concentration of the species $X$.

In a spatially-resolved model, we can decompose the reaction (\ref{eq:A+B=C}) 
in two steps \cite{Agmon90}:
\begin{equation}
A+B\overset{k_{\rm D}}{\underset{k_{\rm D,b}}{\rightleftharpoons}} A\cdot B\overset{k_{\rm a}}{\underset{k_{\rm d}}{\rightleftharpoons}}C.
\end{equation}
In the first step, particles $A$ and $B$ find each other and form an encounter complex $A\cdot B$, which has
not yet reacted to a final product; this occurs via a
diffusion-limited rate $k_{\rm D} = 4 \pi R D$, where $R =
(R_A + R_B)/2$ is the cross section with $R_X$ the
diameter of particle $X$, and $D = D_A + D_B$, with $D_X$ is the
diffusion constant of species $X$ \cite{Agmon90}. 
{\em Given} that the particles are
in contact, the reaction can then proceed according to the intrinsic
reaction rate $k_{\rm a}$. The rates $k_{\rm d}$ and $k_{\rm D,b}$ denote the
intrinsic dissociation rate and the rate at which the particles in the
encounter complex diffuse away into the bulk, respectively
\cite{Agmon90,VanZon06}. It can be shown \cite{Agmon90} that the equilibrium constant is given by 
\begin{equation}
\label{eq:K_eq}
K_{\rm eq} = \frac{k_{\rm a}}{k_{\rm d}} = \frac{k_{\rm on}}{k_{\rm
      off}}
\end{equation}
and that the macroscopic forward and backward rate constants are given
by, respectively,
\begin{eqnarray}
\label{eq:k_f}
\frac{1}{k_{\rm on}} &=&\frac{1}{k_{\rm a}} + \frac{1}{k_{\rm D}},\\
\frac{1}{k_{\rm off}} &=& \frac{1}{k_{\rm d}} + \frac{K_{\rm eq}}{k_{\rm
      D}}.
\label{eq:k_b}
\end{eqnarray}.

We will use Brownian Dynamics to simulate not only the diffusive
motion of the particles, but also the reactions between them. At each
step of the algorithm, each particle is given a trial displacement
according to a distribution that follows from the diffusion equation,
as described below. If the move does not lead to an overlap with
another particle, the move is accepted. Importantly, this procedure
naturally simulates the formation of the encounter complex with a rate
$k_{\rm D}$, provided that the step sizes are smaller than the
diameters of the particles. If two particles are close to each other,
and thus form an encounter complex, a trial displacement of one of the
two can lead to an overlap; this overlap leads to a reaction with a
probability, as derived below, that is consistent with the {\em
  intrinsic} reaction rate $k_{\rm a}$. Conversely, at each step of
the algorithm a product particle $C$ can dissociate with a probability
consistent with the intrinsic dissociation rate $k_{\rm d}$. If a
trial dissociation move is accepted, then the particles $A$ and $B$
have to be put back in the encounter complex. The question, however,
is: at which distance should the particles be put back relative to each
other? The Brownian Dynamics scheme makes an error in the dynamics of
order $\Delta t$. This might suggest that the precise location is not
critically important, as long as the distance is smaller than $\Delta
r \sim \sqrt{D \Delta t}$. However, not every choice obeys detailed
balance, and, as we will show, a choice that does not obey detailed
balance will lead to systematic errors. We now derive the detailed
balance condition.

The detailed-balance condition for {\em one} given pair of particles
$A$ and $B$ is
\begin{equation}\label{eq:DB_space}
P_{\rm unbound} ({\bf r}) d{\bf r} P_{\rm u\to b}=P_{\rm bound}P_{\rm b\to u},
\end{equation}
where $P_{\rm bound}$ is the probability that the two
particles are bound, and $P_{\rm unbound}({\bf r}) d{\bf r}$ is the probability that the particles
$A$ and $B$ are separated by a vector between ${\bf r}$ and ${\bf
  r}+d{\bf r}$. We now first derive
the ratio $P_{\rm
  bound}/(P_{\rm unbound}({\bf r})d{\bf r})$. 
To this end, let us consider the probability $P({\bf r}_A^{N_A},{\bf
r}_B^{N_B},{\bf r}_C^{N_C};\{N_A,N_B,N_C\})d{\bf r}^{N_A}
d{\bf r}^{N_B} d{\bf r}^{N_C}$ that the system has
$(N_A,N_B,N_C)$ molecules {\em and} that these molecules are located
at positions $\{{\bf r}_A^1,\cdots,{\bf r}_A^{N_A}\}$, $\{{\bf
r}_B^1,\cdots,{\bf r}_B^{N_B}\}$, $\{{\bf r}_C^1,\cdots,{\bf
r}_C^{N_C}\}$. This probability is given
by
\begin{eqnarray}
P({\bf r}_A^{N_A},{\bf r}_B^{N_B},{\bf r}_C^{N_C};\{N_A,N_B,N_C\})=
P_N(N_A,N_B,N_C) \times{\cal P}({\bf r}_A^{N_A},{\bf r}_B^{N_B},{\bf
r}_C^{N_C}|\{N_A,N_B,N_C\}),
\end{eqnarray}
where $P_N(N_A,N_B,N_C)$ is the probability that the system has
$(N_A,N_B,N_C)$ molecules and ${\cal P}$ is the {\em conditional}
probability density that a given number $\{N_A,N_B,N_C\}$ of molecules
occupy those particular positions. As discussed in more detail in
Appendix \ref{app:db}, $P_N(N_A,N_B,N_C)$  is given by 
\begin{equation}
P_N(N_A,N_B,N_C) = \frac{q_{A,\rm{cm}}^{N_A} q_{B,\rm{cm}}^{N_B} q_{C,\rm{cm}}^{N_C}V^{N_A+N_B+N_C}}{N_A!N_B!N_C!}\frac{1}{\cal{Q}},
\end{equation}
where $\cal{Q}$ is the partition function of the system.
The conditional probability density
is the probability density of finding $\{N_A,N_B,N_C\}$
indistinguishable ideal particles in a volume $V$:
\begin{equation}\label{eq:thisone}
{\cal P} ({\bf r}_A^{N_A},{\bf r}_B^{N_B},{\bf
r}_C^{N_C}|\{N_A,N_B,N_C\}) = \frac{N_A ! N_B ! N_C
!}{V^{N_A+N_B+N_C}}.
\end{equation}

Combining the above equations, yields the following expression for the
probability density:
\begin{equation} \label{eq:diff}
P({\bf r}_A^{N_A},{\bf r}_B^{N_B},{\bf r}_C^{N_C};\{N_A,N_B,N_C\}) =
\frac{q_{A,\rm cm}^{N_A} q_{B,\rm cm}^{N_B}q_{C,\rm
    cm}^{N_C}}{\cal Q},
\end{equation}
where $q_{X,{\rm cm}}$ is the partition function corresponding to the degrees of freedom od the
center of mass of a particle X, and ${\cal Q}$ is canonical partition function of the system.
The ratio between the probability densities of being in a state after and before the transition is:
\begin{equation} \label{eq:diff2}
\frac{P({\bf r}_A^{N_A-1},{\bf r}_B^{N_B-1},{\bf
r}_C^{N_C+1};\{N_A-1,N_B-1,N_C+1\})}{P({\bf r}_A^{N_A},{\bf
r}_B^{N_B},{\bf r}_C^{N_C};\{N_A,N_B,N_C\})} = \frac{q_{C,\rm
cm}}{q_{A,\rm cm}q_{B,\rm cm}}.
\end{equation}
By taking $N_A=1$, $N_B = 1$,$N_C =0$, we obtain $P_{\rm bound} /(
P_{\rm unbound}({\bf r}) d{\bf r})$:
\begin{equation}\label{eq:DBbis}
\frac{P_{\rm bound}}{P_{\rm unbound}({\bf r})d{\bf r}} =\frac{P({\bf
    r}_C;0,0,1)d{\bf r}_C}{P({\bf r}_A,{\bf
r}_B;1,1,0)d{\bf r}_Ad{\bf r}_B}= \frac{q_{C,\rm
cm}}{q_{A,\rm cm}q_{B,\rm cm}d{\bf r}}=\frac{K_{\rm eq}}{d{\bf r}}.
\end{equation}
 
Using Eq. \ref{eq:K_eq}, the detailed-balance condition, Eq. \ref{eq:DB_space}, then becomes
\begin{equation}
\frac{P_{\rm bound}}{P_{\rm unbound} ({\bf r})d{\bf r}} = \frac{P_{\rm u\to
      b}}{P_{\rm b\to u}d{\bf r}} = \frac{k_{\rm a}}{k_{\rm d}d{\bf
    r}}
\end{equation}

As discussed above, the association between the particles in the
encounter comlpex to form the product $C$ consists of a two-step
process: 1) a ``generation'' move or ``trial'' move, in which an
overlap is generated with a probability $P_{\rm gen,f}$; 2) an
``acceptance'' move, in which the overlap is accepted with probability
$P_{\rm acc,f}$; the product of the probabilities of these moves is
related to the intrinsic reaction rate $k_{\rm a}$. Similarly, the
dissociation move also consists of two steps: 1) a ``trial'' move, in
which the dissociated particles are put at a vector between ${\bf r}$
and ${\bf r}+d{\bf r}$ with
probability $P_{\rm gen,b} ({\bf r})d{\bf r}$; 2) an ``acceptance'' move, in
which the trial move is accepted with probability $P_{\rm acc,b}$; the
product of the probabilities of these moves is related to the
intrinsic dissociation rate constant $k_{\rm d}$. The
detailed-balance condition can thus be written as \cite{Frenkelbook}
\begin{equation}
\label{eq:DB2}
\frac{P_{\rm bound}}{P_{\rm unbound} ( {\bf r})d{\bf r}} = \frac{P_{\rm gen,f}
  ({\bf r}) P_{\rm acc,f}}{P_{\rm gen,b}d{\bf r}({\bf r})P_{\rm acc,
    b}} = \frac{k_{\rm a}}{k_{\rm d}d{\bf r}}.
\end{equation}
This is the principal result of this section. Below, we discuss in
detail how this rule is implemented in our BD scheme. In appendix
\ref{app:db}, we discuss how this detailed-balance rule is related to
the detailed-balance rule for a well-stirred system, where we do not
account for the positions of the particles in space.

\subsection{Simulation scheme}

\noindent It is instructive to consider the association between one
particle $A$ and one particle $B$. We can assume without loss of
generality that $D_A\!=\!0$, {\em i.e.} that the $A$ particle does not
diffuse in the simulation box. It is then convenient to position it at
the center of the box. The single $B$ particle moves by free diffusion
with diffusion coefficient $D_B\equiv D$. At every simulation step,
the system is propagated by a fixed time $\Delta t$.

In the absence of the $A$ particle, the motion of the $B$ particle
is simply described by the Einstein equation:
\begin{equation}\label{eq:diff_part}
\frac{\partial}{\partial t}\,p(\textbf r ^\prime,t+\Delta t|\textbf
r,t)=D\,\nabla ^2p(\textbf r ^\prime,t+\Delta t|\textbf r,t),
\end{equation}
where $p(\textbf r ^\prime,t+\Delta t|\textbf r,t)$ is the
probability of finding the particle at position $\textbf r ^\prime$
at time
$t+\Delta t$, given that it was at $\textbf r$ at time $t$.
We know with certainty the position of the particle at the
initial time. We also impose that at time $t+\Delta t$ the probability of
finding the particle in space vanishes as we move far away from the
initial position ${\bf r}$. We can then formulate the following
boundary conditions for Eq.~(\ref{eq:diff_part}):
\begin{equation}\label{eq:bc1}
p(\textbf r ^\prime,t+\Delta t|\textbf r,t)=\delta(\textbf r^\prime
- \textbf r),
\end{equation}
\begin{equation}\label{eq:bc2}
p(|\textbf r ^\prime|\rightarrow\infty,t+\Delta t|\textbf r,t) =0.
\end{equation}
The solution of (\ref{eq:diff_part}) with conditions (\ref{eq:bc1}) and
(\ref{eq:bc2}) is a Gaussian function, whose variance is proportional to
$\Delta t$:
\begin{equation}
p(\textbf r ^\prime,t+\Delta t|\textbf r,t)=\frac{1}{(2\cdot
2D\Delta t)^{3/2}}\exp\bigg\{-\frac{(\textbf r ^\prime-\textbf r)^2}
{2\cdot 2D\Delta t}\bigg\}.
\end{equation}
This time-dependent probability distribution can be used to generate
new positions for the $B$ particle at every time step $\Delta t$ \cite{NumRec}.

In the presence of the $A$ particle, a
reaction can occur when the $B$ particle overlaps with the $A$
particle. In order to describe the association and dissociation
reactions, we have to specify $P_{\rm gen}({\bf r})$, $P_{\rm acc,f}$,
$P_{\rm gen,b}$, $P_{\rm acc,b}$ in such a way that detaield balance,
Eq. \ref{eq:DB2}, is obeyed. We first discuss $P_{\rm gen,f} ({\bf r})$, then the two quantities related to the backward move, $P_{\rm
  gen,b}({\bf r})$ and $P_{\rm acc,b}$, and then we discuss the
probability by which the trial association move (the overlap) should
be accepted, $P_{\rm acc,f}$.

The quantity $P_{\rm gen,f}({\bf r})$ can be computed analytically:
let us consider the single particle $A$ held fixed in a center of a
large box, whose edges lie far enough to be neglected in the following
derivation. Using a polar reference frame whose origin coincides with
the center of the $A$ sphere, we can compute the probability that a
$B$ particle initially at position $\textbf r$ is displaced to a
position $\textbf r ^\prime\in\Sigma$, where $\Sigma$ is the excluded
volume for $B$ (a sphere, centered in the origin, with radius
$R=R_A\!+\!R_B$):
\begin{equation}\label{eq:g(r)}
p(\textbf r \rightarrow\Sigma)=\int_0^R r^{\prime 2}
dr^\prime\int_0^\pi \sin\theta d\theta\int_0^{2\pi}d\varphi\,
p(\textbf r ^\prime,t+\Delta t|\textbf r,t)\equiv g(r,\Delta t).
\end{equation}
The function $g(r)$ can be computed analytically, is radially symmetric
and depends on the Brownian Dynamics time step $\Delta t$. Details are given
in Appendix \ref{App:g}. We will not indicate anymore the dependence
of various quantities on $\Delta t$, since this parameter is kept
constant during the whole simulation. We remind the reader that in the function $g(r)$, $r$ represents the position from
which the $B$ particle leaves, given that the move led to an overlap with $A$.
We set then $P_{\rm gen,\,f}({\bf r})\!=\!g(r)\Omega(\theta,\varphi)$,
where $\Omega(\theta,\varphi)$ is the uniform angular distribution
on the sphere.

Dissociation is modelled as a first order reaction event, with a
Poissonian distribution of waiting times: $P(t)=k_{\rm
  d}\exp(-k_{\rm d}t)$. The probability that the reaction has not
happened at time $t$ is then $S(t)=1-\int_0^t
P(t^\prime)dt^\prime=\exp(-k_{\rm d}t)$. Therefore, the probability
a reaction {\em does} happen is $1\!-\!\exp(-k_{\rm d}t)\simeq
k_{\rm d}\Delta t$ if $k_{\rm d}\Delta t\ll 1$.  If we choose time
steps $\Delta t$ such that $\Delta t\ll 1/k_{\rm d}$, the
probability that an event happens within $\Delta t$ can then be
approximated to $k_{\rm d}\Delta t$. We therefore accept the
dissociation move with a probability  $P_{\rm acc,\,
  b}=k_{\rm d}\Delta t$.

Once we have determined that a dissociation event has happened, we
must determine a new position for the $B$ particle in the reaction
box. The crux of our BD algorithm is {\em to generate a reverse move
  according to a probability distribution $P_{\rm gen,b} ({\bf r})$ that is the
  same as that by which the forward move is generated, $P_{\rm
    gen,f}({\bf r}) = g(r)\Omega (\theta,\phi)$, but properly
renormalised}. The normalisation factor can be obtained by integrating
$P_{\rm gen,f}({\bf r})$ over all initial distances $r$:
\begin{equation}
\int_R^\infty dr\int d\Omega\;\Omega(\theta,\varphi) g(r,\Delta
t)r^2=4\pi I(\Delta t)\ll V,
\label{eq:4piI}
\end{equation}
where $I\!=\!\int_R^\infty g(r)r^2 dr$. During a dissociation move, the particle is thus put at a
vector between ${\bf r}$ and ${\bf r}+d{\bf r}$ according to $P_{\rm
  gen, b}({\bf r})d{\bf r} =\frac {d{\bf r}} {4\pi I} g(r)\Omega(\theta,\varphi)$.

Using Eq. \ref{eq:DB2}, we can now obtain the
desired acceptance probability for the forward move:
\begin{eqnarray}\label{eq:Pacc1}
P_{\rm acc, f} &=& \frac{P_{\rm bound}}{P_{\rm unbound}}\;\frac{P_{\rm gen, b}}{P_{\rm gen, f}}P_{\rm acc, b}\nonumber \\
           &=& \frac{k_{\rm a}}{k_{\rm d}d{\bf r}}\;\frac{g(r)\Omega(\theta,\varphi)d\bf r}{g(r)\Omega(\theta,\varphi)\,4\pi I}k_{\rm d}\Delta t\nonumber \\
           &=& \frac{k_{\rm a}\Delta t}{4\pi I}.
\end{eqnarray}
The above expression has a meaningful interpretation: the intrinsic
association rate $k_{\rm a}$ can be written as the product of two
factors: 1) a collision frequency $4\pi I / \Delta t$ and 2) the
probability $P_{\rm acc,f}$ that a collision leads to a reaction. The
dominant contributions to the integral $I$ come from distances $r$
that are short compared to $\sqrt{D \Delta t}$ (see Eqs. \ref{eq:g(r)}
and \ref{eq:4piI}). In the limit that the time step $\Delta t  \to
0$, the rate $k_{\rm a}$ should thus approach the intrinsic
association rate $k_{\rm a}$ as used in theories of
diffusion-influenced reactions \cite{Agmon90}; here the intrinsic
association rate $k_{\rm a}$ is defined as the association rate given
that the particles $A$ and $B$ are in contact. We also note that this
is the intrinsic association rate as used in Green's Function Reaction
Dynamics \cite{VanZon05,VanZon05_2}.

\subsection{Algorithm outline}

Let us consider a system with $M$ particles of type $B$ and one particle
of type $A$, held fixed at the center of box of volume $V$. For
convenience, we choose as initial state the situation in which there is
no bound state $C$. 

\begin{enumerate}
\item Generate an initial position for the $B$ particles in the available volume.

\item Select randomly one of the particles among species $B$ and $C$.

\item\begin{enumerate}
 \item If the particle is type $B$, for each cartesian coordinate, generate a new position according to a
  Gaussian distribution with zero mean and standard deviation
  $\sqrt{2D\Delta t}$: $x_{\rm new}=x_{\rm old}\!+\!N(0,\sqrt{2D\Delta t})$,
  where $\Delta t$ the Brownian Dynamics time step. 

  \item If the displacement move leads to an overlap of the $B$ particle with $A$, that
  is if $|\textbf{r}_A-\textbf{r}_B|<R_A+R_B$, attempt a reaction
  according to a probability $P_{\rm acc,\,f}=k_{\rm a}\Delta t/(4\pi I)$.

  \item If the trial reaction move is accepted, remove the $B$ particle from the box,
  and substitute the $A$ particle with a $C$. This new particle is
  not diffusing in the box.

  \item If the trial reaction move is rejected, put the $B$ particle back to its original position.
  \end{enumerate}

\item\begin{enumerate}
  \item If the particle is type $C$, try a backward reaction with
  probability $P_{\rm acc,\, b}=k_{\rm d}\Delta t$.

  \item If the trial reaction move is accepted, substitute the $C$ particle with an
  $A$ particle, create a new $B$ particle whose radial position is
  drawn from the normalised distribution $g(r)/I$ and the angular position
  from the uniform distribution $\Omega(\theta,\varphi)$. If this leads to 
  an overlap with another $B$ particle, reject the move.

  \item If the trial reaction move is rejected, keep the identities and positions of the
  particles.
\end{enumerate}

\item Repeat step 2. and 3. or 4. $M$ times, then increase the simulation
time by $\Delta t$.

\end{enumerate}

Keeping particle $A$ and $C$ fixed could mimic for example a system
where one reactant is anchored to some rigid scaffold. A relevant
biological example is the binding of proteins to DNA in a bacterial
cell, particle $A$ representing a binding site on the DNA, typically
in proximity of some gene. In this case, the motion of $A$ is only
related to the fluctuations of the polymer, which happen on time
scales much longer than the diffusion of proteins in the bacterial
cytoplasm, and can therefore be neglected.  The scheme could
straightforwardly be extended to the situation in which the $A$
particle also moves, or cases with more reaction channels.

\section{Tests}\label{sec:tests}

In this Section, we check the BD scheme by comparing the simulation
results with analytical results.  Our scheme was built upon a series
of assumptions, which should all be satisfied simultaneously. In
particular, 1) the time steps should not be too large, as a BD
algorithm is not able to resolve the system at time scales below the
time step $\Delta t$; 2) the acceptance probabilities for the forward
and backward reactions should be small ($P_{\rm acc,f}\!\ll\! 1$,
$P_{\rm acc,b}\!\ll\! 1$). In particular, the algorithm does not
resolve the precise moment in time when the association and
dissociation events happen. Dynamical quantities could therefore
exhibit systematic errors, which must vanish in the limit $\Delta
t\!\to\! 0$.  On the other hand, on long enough time scales, even
dynamical poperties should be reproduced, provided that the conditions
listed above are obeyed.  In the case of two particles, the
probability distribution $p({\bf r},t|{\bf r}_0)$ of finding the
particles separated by a vector ${\bf r}$ at time $t$ given that
initially they were separated by ${\bf r}_0$, as well as their
survival probability $S(t|{\bf r}_0)$ \cite{Agmon90}, have been
computed analytically \cite{Kim98}; they will provide a stringent test
for the {\em dynamics} of our scheme. As our algorithm obeys detailed
balance, {\em equilibrium} quantities, such as the average time spent
in the bound state, must be correctly reproduced {\em for all time
  steps $\Delta t$}.

\subsection{Irreversible Reactions}\label{subsec:irr_test}

We begin by simulating the {\em irreversible} reaction $A+B
\overset{k_{\rm a}}{\longrightarrow} C$, within the following setup:
a single particle $A$ is held fixed in an unbounded system, and a single particle $B$ is positioned on a spherical surface at an initial
distance $r_0$ from $A$, with a random angle. The particles have the same radius:
$R_A\!=\!R_B\!=\!R/2$. We run the algorithm for a time $t_{\rm sim}$,
and we record the final radial position of the particle $B$. In the case that
a reactive event happens before $t_{\rm sim}$, we stop the run. 
After a large number of runs, we collect
the final positions of the $B$ particle in an histogram, 
normalised to the fraction of $B$ particles which have survived 
until the final time. This histogram should reproduce the irreversible probability distribution $p_{\rm irr}(r,t_{\rm
sim}|r_0,0)$ \cite{Kim98}. This quantity represents the probability of finding the two particles
at time $t_{\rm sim}$ separated by a distance $r$, given an initial
separation of $r_0$ at $t_0\!=\!0$. 
We note that this probability distribution is not normalised to unity:
the integral over space of $p_{\rm irr}$ is the survival probability 
of the particle, which is the probability that the particle has not reacted at the final time. 
Formally:
\begin{equation}
4\pi \int_R^\infty p_{\rm irr}(r,t|r_0)\,r^2\, dr=S_{\rm irr}(t|r_0).
\end{equation}
We are thus able to simultaneously test our algorithm twice:
comparing the analytical curve with the profile of our histogram, and the
area of the histogram with the analytical value of the survival probability.

Results are collected in Figure~\ref{fig:pirr}: we simulate the
irreversible reaction for 4 different simulation times, from $t_{\rm
  sim}\!=\!10^{-4}\tau$ to $t_{\rm sim}\!=\!10^{-1}\tau$, where
$\tau\!=\!R^2/D$ is the natural time scale of the system.  Particles
are initially positioned at contact: $r_0\!=\!R$.  We set the time
step $\Delta t=10^{-4}t_{\rm sim}$, which corresponds to $P_{\rm
  acc,f}< 0.14$. It is seen that both the shape and the area of the
irreversible probability disitribution function is correctly captured
by our algorithm.  In the case of $t_{\rm sim}\!=\!10^{-1}\tau$,
however, we needed to use $\Delta t=10^{-5}t_{\rm sim}$. 
In the Inset of Fig.~\ref{fig:pirr}, we show that in this last case,
larger time steps lead our BD scheme to underestimate the survival
probability. The deviation from the analytical results for large
$\Delta t$ is due to the interplay bewteen a number of
assumptions. One is that $1\!-\!\exp(P_{\rm acc,f})\simeq P_{\rm
  acc,f}$. A more important factor is that we compare the numerical
results against analytical results of an analysis in which $k_{\rm
  on}$ corresponds to the intrinsic association rate for two particles
that are {\em at contact} \cite{Kim98}, while in our scheme the particles can
already react when they are separated by a distance $\sim \sqrt{D
  \Delta t}$. This overestimates the number of
reactions and hence decreases the survival probablity, consistent with
the results shown in the inset of Fig. \ref{fig:pirr}. Another way of
putting this that in our BD algorithm, the intrinsic association rate
is higher than that used in the analytical calculations.

\subsection{Reversible Reactions}\label{subsec:gill_comp}

We extend now the dynamical test performed above to the case of the
{\em reversible} reaction $A+B\overset{k_{\rm a}}{\underset{k_{\rm
      d}}{\rightleftharpoons}} C$.  An analytical solution to the
problem, $p_{\rm rev}(r,t|r_0,t_0)$ is known for one A and one B
particle \cite{Kim98}.  In this test, we adopt the same setup and a
similar procedure as for the irreversible case, except that we do not
stop the run after a reaction, but we let the particle dissociate. At
$t\!=\!t_{\rm{sim}}$ we check whether or not the $B$ particle is in
the bound state. If it is not, we record the final position. The
histogram of final positions of the $B$ particles is normalised to the
number of survivors at $t\!=\!t_{\rm{sim}}$, and compared with the
analytical curve $p_{\rm rev}(r,t_{\rm sim}|r_0,0)$ \cite{Kim98}. The
fraction of runs ending in the unbound state yields an estimate for
the survival probablity $S_{\rm rev}(t_{\rm sim}|r_0)$.  Again, we
initially put the $B$ particle at contact ($r_0\!=\!R$), so that a
larger number reactions and dissociations can happen within
$t_{\rm{sim}}$. This choice will provide a stringent test for the
dynamics of the system. The parameters of the system are the same as
in Section \ref{subsec:irr_test}, with the addition of the
dissociation rate $k_{\rm d}\!=\!1\tau^{-1}$. 
Similar results were
obtained for larger values of $k_{\rm d}$, as well as for other
values of $r_0$, $D$, and $k_{\rm a}$.

In Figure~\ref{fig:prev}, we plot $p_{\rm rev}(r,t_{\rm sim}|R,0)$ for
$t_{\rm{sim}}$ ranging from $10^{-1}\tau$ to $10^{-4}\tau$, with the
same time steps as in the previous test.  We find the BD algorithm
correctly reproduces both the shape and the area of the analytical
probability distribution.  Similarly to Figure~\ref{fig:pirr}, we show
in the Inset $p_{\rm rev}$, computed for $t_{\rm
  sim}\!=\!10^{-1}\tau$, for decreasing time steps $\Delta t$: as
expected on the basis of the results for the irreversible reaction
(Fig. \ref{fig:pirr}), simulations with large time steps
underestimate the survival probability.  

For the next tests, we consider a setup similar to that used above,
but we allow for multiple $B$ particles, and we enclose our system in
a cubic simulation box of volume $V$, endowed with reflecting walls.
All the $B$ particles can bind to the single $A$ particle, and do not
interact among themselves. This last assumption, valid only in the
limit of low packing fractions $\phi$, is satisfied in our simulations
where $\phi\!<\!0.02$. Conversely, particles $B$ and $C$ interact as
hard spheres, {\em i.e.} they are not allowed to overlap. 
We investigate whether {\em equilibrium} properties of the system,
such as the probability of being in the bound state $C$ ($p_{\rm
  bound}$), are correctly reproduced by the BD algorithm.
The probability
$p_{\rm bound}$ can be evaluated by measuring the time when the $C$
particle is present in the system, with respect to the total
simulation time. The mean field value for this quantity can be
obtained from the macroscopic rate equation in steady state:
\begin{equation}\label{eq:p_bound}
p_{\rm{bound}}=\frac{K_{\rm{eq}}N_B}{K_{\rm{eq}}N_B+V^*}\;,
\end{equation}
where $K_{\rm{eq}}\!=\!k_{\rm a}/k_{\rm d}$, and $V^*=V\!-\!\frac 4
3 \pi(R_A+R_B)^3$.     
We note here that this prediction should be valid when the radial
distribution at contact equals unity; given the low density of
particles in our system this should be the case.

We simulate the system with a varying number $N_B$ of $B$ particles,
with a fixed time step $\Delta t\!=\!10^{-4}\tau$.  We choose $K_{\rm
  eq}\!=\!V$, so that $p_{\rm bound}(N_B\!=\!1)=0.5$.  The enclosing
box measures $(20R\times 20R\times 20R)$.  Figure~\ref{fig:p_bound_NB}
compares the results of our simulations with the prediction of
Eq. \ref{eq:p_bound}: we see a clear agreement.
To illustrate that obeying the detailed-balance condition is
important, we also performed a series of simulations in which the
particles after dissociation were put at contact; in other words, we
considered a function $g(r)=\delta(r\!-\!R)$. This move {\em does}
violate detailed balance and indeed affects a correct estimate for
$p_{\rm bound}$, as shown in Inset A of Figure~\ref{fig:p_bound_NB}:
the incorrect procedure overestimates the time the particle spends in
the bound state, especially for a low number of $B$ particles. These data
clearly show that a naive treatment of the dissociation events leads
to systematic errors, which are especially severe when the system has
a low number of reactants, as it is often the case in
biochemical networks.  Finally, we tested whether the equilibrium
properties of the system do not depend on the chosen time step.  To
this end, we compute $p_{\rm bound}$ for $N_B\!=\!1$ and different
values of the time step $\Delta t$.  As illustrated in the Inset B of
Figure~\ref{fig:p_bound_NB}, we obtain a good agreement even for very
large time steps, where probably the {\em dynamics} of the system is
not entirely natural. 

Finally, we compare our Brownian Dynamics algorithm with the
Stochastic Simulation Algorithm, based on a Kinetic Monte Carlo scheme
that propagates the system according to the solution of its
zero-dimensional chemical master equation \cite{Gillespie77}. This
scheme accounts only for the stochasticity arising from the
fluctuations in the number of particles; spatial fluctuations due to
the diffusive motion of particles are completely neglected. The system
is thus assumed to be well-stirred at all times.  We consider the
reversible reaction $A+B\rightleftharpoons C$ for $N_B\!=\!1$: in the
SSA, the association times follow a Poisson distribution, with mean
$1/k_{\rm f}$, where $k_{\rm f}^{-1}=1/(4\pi D R)\!+\!k_{\rm a}^{-1}$
is the macroscopic forward rate.

We collect the association times for a BD run with $V\!=\!64000
R^3,\Delta t\!=\!10^{-4}\tau,k_{\rm a}\!=\!100R^3/\tau,k_{\rm
  off}\!=\!1000\tau^{-1}$, and we compare it with an SSA run obtained
with the same set of parameters, but using the modified association
rate $k_{\rm f}$.
Figure~\ref{fig:treact} compares the two distributions: the BD line
shows a marked increase in the number of association events at short
times, as compared to the Poissonian distribution with mean $k_{\rm
  f}$ of the SSA.  This effect has a purely spatial origin and has been
previously observed (\cite{VanZon05_2,VanZon06}): when particles
dissociate in space, their distance is still very small, therefore the
probability of an immediate rebinding in next few times steps is very
high. Long association times, in a BD simulation, are related to
particles which have wandered diffusively in the box, and have finally
found the target. The distribution of such times is again exponential,
with a constant $k_{\rm f}$.  

The test above show that our BD algorithm, which rigorously obeys
detailed balance, 
correctly reproduces the equilibrium properties and provides a good
description of the dynamics of the system in time and space.

\section{Application: the push-pull model}\label{sec:p-p}

In this Section, we apply our Brownian Dynamics to a simple model of a
push-pull network. In this network, two antagonistic enzymes continually
covalently modify and demodify a substrate, respectively; a well-known
example is a protein that is phosphorylated and dephosphorylated by a
kinase and a phosphatase, respectively. The first enzyme converts a substrate
molecule into an ``active'' state: bearing in mind the phosphorylation
example, we call this active substrate $S_p$, and the enzyme $K$
(kinase). A molecule in the active state $S_p$ can be brought back to
the original state $S$ under the action of a second enzyme, $P$
(phosphatase). The model is nicknamed ``push-pull'', as the substrates
are continuously switching between the two states, while consuming
energy. The reactions with the enzymes are described according to
Michaelis-Menten kinetics: the two reactants form first an
intermediate bound state, which can lead either to a dissociation or
to the release of a converted molecule. In \cite{Goldbeter81}, the
model is solved at the level of the Macroscopical Rate Equation at
steady state, which yields the average behavior of the system.

Goldbeter and Koshland showed that such a system can display an {\em
  ultrasensitive} behavior (that is, a sensitivity curve steeper than
the conventional response showed by the Michaelis-Menten mechanism)
without the need of introducing cooperative interactions
\cite{Goldbeter81}. More precisely, the interplay between two
converter enzymes operating in opposite directions on a target whose
quantity is conserved can give rise to a switch-like response in the
steady-state fraction of modified molecules, when the ratio between the
conversion rates is varied. The requirement for such a sharp
transition is the saturation of the enzymes: the effective conversion
rates then become independent on the number of substrate molecules,
thus making the reaction rates ``zero-order'' in the substrate concentration.

The above-mentioned analysis does not however account for any kind of
fluctuations that may arise from the low number of reactants, the
stochastic behavior of the chemical reactions, or the diffusion of the
molecules in space. In \cite{Berg00}, the same model is studied at the
level of the chemical master equation, taking into account finite-size
effects that arise in real systems, that is the discreteness and the
possible low copy number of enzymes and substrates. In order to
achieve ultrasensitivity, the enzymes must be saturated, and therefore
their concentration is likely to be very low. Large fluctuations are
then observed around their average behavior: the authors show that the
results obtained with a mesoscopic approach reduce to those of the
macroscopic analysis of \cite{Goldbeter81} {\em only} when the number
of molecules is sufficiently large. If this is not the case, as it can
easily happen in a bacterial cell where some species are present only
in few dozens of copies, the sensitivity of the system is reduced, and
the response is less steep than the macroscopic theory would predict.
This deviation can be easily understood when one realises that high
sensitivity corresponds to highly saturated enzymes. In this regime,
the reaction rates do not depend on the number of substrate
molecules. The system
performs a random walk in the number of $S$ molecules and it is thus
subject to large fluctuations. Our Brownian Dynamics algorithm now
allows us to study the effect of spatial fluctuations due to the
diffusive motion of the molecules.

The system we consider is defined by the following
set of reactions:
\begin{subequations}
\label{eq:push-pull}
\begin{align}
\mathrm{Reaction} & \quad &\mathrm{Rate} \nonumber\\
\hline
S+K\rightleftharpoons KS & \quad & k_{\rm on1},\:k_{\rm off1} \\
KS \to K+S_p & \quad & k_1 \\
S_p+P\rightleftharpoons PS_p & \quad & k_{\rm on2},\:k_{\rm off2} \\
PS_p \to P+S & \quad & k_2 .  
\end{align}
\end{subequations}
Here $k_{\rm f}$ stands for the macroscopical association rate. The
system will be simulated with the BD algorithm in a rectangular box of
dimensions $x_{\rm box}\!=\!20R,y_{\rm box}\!=\!10R, z_{\rm
  box}\!=\!10R$, with a single kinase and a single phosphatase
molecule, held fixed at distance $\Delta\!=\!0.5x_{\rm box}$ on the
central axis of the box, as depicted in Figure~\ref{fig:snapshot}. The
system is initially prepared with $N_{S\rm tot}$ particles,
distributed in the two states according to the solution of the
macroscopical rate equation. In the following, we investigate the
effect of spatial fluctuations of the substrate molecules on the
input-output relation of the system, and we compare the BD results to
those obtained with the mean-field and the
zero-dimensional chemical master equation approach.

The input-output relation is defined as the mean fraction of
phosphorylated substrate molecules $\langle S_p\rangle/N_{S\rm tot}$
as a function of the ratio $k_1/k_2$. We compute it with 80 substrate
molecules in the simulation box, in order to meet the requirement
$N_S\!\gg\!N_K$ and $N_{S_p}\!\gg\!N_P$ $(N_S\!+\!N_{S_p}=N_{S\rm
tot})$. The parameters governing the steepness of the sigmoid curves
are the Michaelis-Menten equilibrium rates $K_1\!=\!(k_{\rm
off1}+k_1)/k_{\rm on1}$ and $K_2\!=\!(k_{\rm off2}+k_2)/k_{\rm
on2}$. In all our simulations we set $K_1\!=\!K_2\!=\!K_{\rm
M}$. In our simulations, 
diffusion constants are in the order of $10^{-4}-10^{-5} R^2/\Delta t$, intrinsic 
association rates between 0.001 and 0.006 $R^3/\Delta t$, dissociation rates vary
between $5\cdot 10^{-4}$ and $5\cdot 10^{-6} (\Delta t)^{-1}$, and production rates are
chosen in the range $10^{-4}-10^{-8} (\Delta t)^{-1}$. To vary $k_1/k_2$ keeping $K_{\rm
M}$ constant, we vary $k_{\rm off1}$ and $k_{\rm 1}$ together, keeping their sum constant.
When $K_{\rm M}/[S_{\rm tot}]\ll 1$ the enzymes are totally saturated and the
change in the fraction of modified proteins is abrupt; on the other
hand, when $K_{\rm M}/[S_{\rm tot}]\geq1$, the rise of the curve becomes
asymptotically close to the hyperbolic Michaelis-Menten shape.

Figure~\ref{fig:sensitivity} shows several examples of the
input-output relation, obtained with three different methods: with the
analytic macroscopic approach of \cite{Goldbeter81} (solid lines),
with SSA simulations as in \cite{Berg00} (diamonds) and with the
Brownian Dynamics algorithm (circles). The BD simulations are performed
with intrinsic association and dissociation rates $k_{\rm a}$ and
$k_{\rm d}$, respectively. In the mean-field analysis and the SSA
simulations of the zero-dimensional chemical master equation, for the
{\em association} reaction the rate constant was chosen to be
that of the macroscopic
association rate $k_{\rm on}$, as given by $k_{\rm on}=(1/k_{\rm
a}\!+\!1/k_{\rm D})^{-1}$, with $k_{\rm a}$ being the intrinsic association rate, and 
$k_{\rm D}=4\pi R D$ the
diffusion-limited rate (see Eq. \ref{eq:k_f}).
For the rate constant of the {\em backward} reaction in the mean-field
analysis and the SSA simulations, we chose the intrinsic reaction rate
and not the macroscopic one given by Eq. \ref{eq:k_b}. The reason is that
the macroscopic dissociation rate takes into account that, upon
dissociation, the dissociated species rebind a number of times before
they diffuse away from each other into the bulk \cite{VanZon06}. As we
have recently shown, association and dissociation reactions can be
described with effective rates given by Eqs. \ref{eq:k_f} and
\ref{eq:k_b} when the associated species can only dissociate, thus
when there is no competing decay channel for the associated species
\cite{VanZon06}. In particular, in Ref. \cite{VanZon06} we studied the
effect of spatial fluctuations due to the diffusive motion of
repressor molecules on the noise in the expression of a gene; the
simulation results showed that the stochasticity in the binding of the
repressor to the DNA resulting from the spatial fluctuations of the
repressor molecules can be a major source of noise in gene expression;
however, this result could be described by renormalising the intrinsic
association and dissociation rates for repressor-DNA binding using the
expressions of Eq. \ref{eq:k_f} and Eq. \ref{eq:k_b}. Here, the
situation is markedly different. The reason is that the associated
species, $KS$ and $PS_p$, can either dissociate or lead to a chemical
modification reaction, upon which the product must diffuse to the
other enzyme in order to be demodified. These reaction channels
compete with one another, and if they compete with one another on the
same time scale, the effective dissociation rate is difficult to
determine. In our system, however, the number of rebindings is in the order of unity.
Therefore, renormalising the rate constants does not substantially change the actual values of the
rate constants. We carried out simulations where we chose either to renormalise both the
association and dissociation rates or neither of them, and we found analogous results. In the
following, we show only BD data obtained with the intrinsic dissociation and association rate.
$K_{\rm M}$ is computed with the intrinsic dissociation rates, and and with $k_{\rm
on}=(k_{\rm D}^{-1}+k_{\rm a}^{-1})^{-1}$.

The different panels of Figure~\ref{fig:sensitivity} show the data for
increasing $K_{\rm M}/[S_{\rm tot}]$. In panel D, $K_{\rm M}/[S_{\rm tot}]\!=\!3.7$ and the
response of the system is very similar to a Michaelis-Menten
kinetics. In this case, the results of the simulations, obtained both
with BD and the SSA, perfectly follow the analytical curve. When
$K_{\rm M}\!>\!1$, both reactions are in the first-order regime, which
means that their rates are proportional to the number of substrate
molecules. As a result, when this number changes, the rates of
conversion in the opposite direction changes immediately. This
counteracts the modification and reduces the effect of
fluctuations. However, as we decrease $K_{\rm M}$, the numerical data
start to deviate from the predicted macroscopic behaviour. In the case
of SSA, the deviation is mild, and barely visible in panel A, where
$K_{\rm M}/[S_{\rm tot}]\!=\!0.05$. In contrast, the BD simulations yield a much more
marked deviation, clearly seen already from panel C, where $K_{\rm M}/[S_{\rm tot}]\!=\!0.52$.

The data in Figure~\ref{fig:sensitivity} confirm and extend the
findings of Ref.~\cite{Berg00}: stochastic fluctuations of the
system dampen the ultrasensitivity, which could be obtained only in an
infinitely large, well-stirred system. The SSA correctly accounts for
the temporal fluctuations arising from the stochastic behavior of
chemical reactions, and for the discreteness of the components, but it
does assume a well-stirred system, where spatial fluctuations can be
neglected. These hypothesis result in a deviation on the order of few
per cents in the ultrasensitive regime. Brownian Dynamics, on the
contrary, properly accounts for temporal {\em and} spatial
fluctuations. These last are related to the diffusive motion of the
substrate molecules; when the concentrations of the species are low,
they become a serious limiting factor which notably reduces the
sensitivity of system. Brownian Dynamics is therefore able to show
that the response of the system can be much less sensitive than
predicted at the macroscopic level in the ultrasensitive regime, if
the species move slowly enough in space,
{\em i.e.} when the system is not well-stirred.

Finally, we emphasise that Brownian Dynamics algorithms can be used to
directly measure spatial properties of the system. Among several
possibilities, we choose to show in Figure~\ref{fig:grad} the spatial
density of particles along the main axis of the box. Data are obtained
for two different values of the Michaelis-Menten constant: $K_{\rm M}/[S_{\rm tot}]\!=\!4.7$ 
($D=10^{-4}R^2/\Delta t, k{\rm on}=0.007R^3/\Delta t,k_1+k_{\rm off1}=2\cdot 10^{-4}(\Delta t)^{-1},N_{S_{\rm tot}}=80$), 
where the system is in the first-order regime, and
$K_{\rm M}/[S_{\rm tot}]\!=\!0.22$ 
($D=10^{-3}R^2/\Delta t, k_{\rm on}=0.01R^3/\Delta t,k_1+k_{\rm off1}=5\cdot 10^{-5}(\Delta t)^{-1},N_{S_{\rm tot}}=80$), 
corresponding to the ultrasensitive regime.
Substrate molecules are very often bound to the enzymes: at the kinase
enzyme, the $S$ density has a sharp peak, and so does the $S_p$
density at the location of the phosphatase enzyme. The height of the
peak is bigger when the system is simulated for a low value of $K_{\rm
  M}$, where enzymes are completely saturated. Interestingly, the
concentration of both molecules shows a gradient along the $x$
direction, higher in the half box where the molecules are
produced. Such gradient is not particularly appreciable for low values
of $K_{\rm M}$: in this regime, enzymes are saturated and dissociation
events are rare. The particles have therefore the time to spread in
the box and reach an homogeneous concentration only occasionally
perturbed by a production event. For high values of $K_{\rm M}$,
particles are produced more often and they do not have the time to
stir in the box, leading to an accumulation close to the production
sites.  The profiles for the two species are completely symmetric, as
these simulations are obtained for $k_1\!=\!k_2$. Naturally, such a spatial
property could not be captured if the system is simulated at the level
of the zero-dimensional chemical master equation.

\section{Discussion}
In this manuscript, we have presented a Brownian Dynamics algorithm
that rigorously obeys detailed balance for equilibrium
reactions. Consequently, the equilibrium properties of biochemical
networks, such as promoter and receptor occupancies, are reproduced
exactly to within the statistical error. Moreover, the association and
dissociation reaction moves are constructed such that they allow for a
meaningful interpretation: as the time step $\Delta t\to 0$, the
association and dissociation rates approach the intrinsic values
corresponding to the reaction rates of the species {\em at
  contact}. This is useful, also because it allows the BD results to
be compared to theoretical results on diffusion-influenced reactions,
which describe reactions as reactions between species at contact
\cite{Agmon90}. The results shown in Figs. \ref{fig:pirr} and
\ref{fig:prev} show that as $\Delta t\to 0$, the BD simulations
correctly describe not on the equilibrium properties, but also the
dynamical properties of a bimolecular reaction.  For larger time
steps, the BD results deviate from the analytical results, but this is
to be expected since the analytical results assume that the molecules
move by diffusion up to the smallest length and time scales, and that
reactions only occur once the molecules have moved by diffusion into
contact. We believe that while the BD results and the
analytical results match for $\Delta t < 10^{-6} R^2/D$
(Figs. \ref{fig:pirr} and \ref{fig:prev}), the BD algorithm gives a
good description of the dynamics over a large range of time steps,
i.e. for $\Delta t < 10^{-4}$, because in
this range the BD results can be fitted to the analytical results with
a different $k_{\rm a}$ and $k_{\rm d}$ (data not shown).

As an illustrative example, we have applied our BD scheme to a model
representing the dynamics of a substrate molecule under the action of
two antagonistic enzymes. This model was previously analysed with
deterministic methods \cite{Goldbeter81}, which revealed an
ultrasensitive behavior in the response of the system when the enzymes
are fully saturated. A study conducted at the level of the Chemical
Master Equation \cite{Berg00}, thus accounting for the low copy number
of the substrate molecules, highlighted that the ultrasensitivity
predicted in Ref.~\cite{Goldbeter81} cannot be achieved when the
concentration of the substrate is very low. Temporal fluctuations
limit then the sensitivity of the system in the ultrasensitive
regime. We repeated the analysis of Ref.~\cite{Berg00} simulating the
system with the SSA, and confirmed their findings.  Furthermore, we
have investigated the role of spatial fluctuations on the system with
BD simulations. Our analysis shows that the sensitivity of the
response curve in the ultrasensitive regime is furtherly reduced when
the diffusion of the reactants is taken into account explicitely. In
particular, when the diffusion of particles is slow and the system is far
from well-stirred, spatial fluctuations are the dominant source of
noise, and the reduction in the gain is significant.

\appendix

\section{Detailed balance for a well-stirred model}\label{app:db}

In the case of the well-stirred model we used in Sec. \ref{subsec:gill_comp} and
\ref{sec:p-p}, the detailed-balance condition is simpler than in the spatially resolved
model. Let $N_A,N_B,N_C$ be the number of $A$, $B$ and $C$ molecules and $V$ the
volume of the system. The configurational partition function of the
system can be written as the following sum of terms in the canonical
ensemble:
\begin{equation}
{\cal Q} = \sum_{\{N\}} Q(N_A,N_B,N_C),
\end{equation}
where $\{N\}$ denotes all possible combinations of $\{N_A,N_B,N_C\}$;
note that we integrated here over the momenta.
The choice of the canonical ensemble is motivated by the assumption that the cell is a closed system, and it 
does not exchange particles with the environment. 

Let us consider the case where $\{A,B,C\}$ are
ideal particles in a volume $V$, except for the fact, of course, that $A$ and $B$ can form $C$.
The configurational integral $Q$ for $\{N_A,N_B,N_C\}$ particles is then:
\begin{eqnarray}
Q(N_A,N_B,N_C) &=& \frac{q_A^{N_A} q_B^{N_B} q_C^{N_C}}{N_A ! N_B !
N_C !}\\
&=&\frac{q_{A,\rm{cm}}^{N_A} q_{B,\rm{cm}}^{N_B} q_{C,\rm{cm}}^{N_C}}{N_A!N_B!N_C!}V^{N_A+N_B+N_C},\nonumber
\end{eqnarray}
where $q_A$ is the molecular partition function for an $A$ particle,
and the factor $1/(N_A!)$ takes into account the indistinguishability of the $A$
particles. The molecular partition function is given by
$q_A\!=\!q_A^{\rm id}q_{A, \rm{cm}}$, where $q_{A,\rm{cm}}$ is the
partition function corresponding to the internal degrees of freedom
relative to the center of mass and $q_A^{\rm id}\!=\!V$ is
the partition function associated with the translational degrees of
freedom of the center of mass. The
probability that the system has $\{N_A,N_B,N_C\}$ molecules,
$P(N_A,N_B,N_C)$, is then:
\begin{equation}
P(N_A,N_B,N_C) = Q(N_A,N_B,N_C)/{\cal Q}.
\end{equation}

Let us now consider the transition from $\{N_A,N_B,N_C\}$
to $\{N_A-1,N_B-1,N_C+1\}$ molecules. The ratio between the
probabilities of being in the state after and before the transition is:
\begin{eqnarray}
\label{eq:p1} 
\frac{P(N_A-1,N_B-1,N_C+1)}{P(N_A,N_B,N_C)} &=&
\frac{N_A N_B}{N_C+1}\frac{1}{V}
\frac{q_{C,\rm cm}}{q_{A,\rm cm}q_{B,\rm
    cm}}. \\
&=&\frac{N_A N_B}{N_C+1}\frac{1}{V} K_{\rm eq} = \frac{N_A N_B}{N_C+1}\frac{1}{V} \frac{k_{\rm f}}{k_{\rm b}}.\label{eq:p1bis}
\end{eqnarray}
Please note that $K_{\rm eq}$ has dimension of volume, such that the
expression on the right-hand-side is indeed dimensionless. The above
expression serves to illustrate the detailed-balance rule
\cite{Frenkelbook}, which states that
\begin{equation}\label{eq:DB_nospace}
P_{\rm unbound}P_{\rm u\to b}=P_{\rm bound}P_{\rm b\to u}.
\end{equation}
Here $P_{\rm unbound}$ is the probability of being in the state $\{N_A,N_B,N_C\}$, $P_{\rm u\to b}$ is the probability of a transition from
$\{N_A,N_B,N_C\}$ to $\{N_A-1,N_B-1,N_C+1\}$, $P_{\rm b\to u}$ is the probability of the reverse move, and $P_{\rm bound}$ is the probability of
being in the state $\{N_A-1,N_B-1,N_C+1\}$. Using Eq. (\ref{eq:p1bis}) and the former relation we obtain:
\begin{equation}
\label{eq:prob_int} 
P_{\rm u\to b}=\frac{k_{\rm f}}{V} N_AN_B\quad {\rm and} \quad P_{\rm b\to u}=k_{\rm b}(N_C+1).
\end{equation}
These transition probabilities precisely correspond
to those used in Monte Carlo simulations of the
zero-dimensional chemical master equation \cite{Gillespie77}.

\section{Derivation of $g(r)$}\label{App:g}

The function $g(r)$, described in Section \ref{sec:methods_BD} is given by the integral of equation (\ref{eq:g(r)}):
\begin{equation}
g(r)=\frac{1}{(\pi\sigma^2)^{3/2}}\int_0^R r^{\prime 2}
dr^\prime\int_0^\pi \sin\theta d\theta\int_0^{2\pi}d\varphi\,
\exp\left(-\frac{r^{\prime 2}-2rr^\prime\cos\theta+r^2}{\sigma^2}\right),
\end{equation}
where $\sigma^2=4D\Delta t$.\\
Elementary methods can be used: integration over the angular
variables yields
\begin{equation}
g(r)=\frac{1}{\sqrt{\pi\sigma^2}}\frac{\exp{\left(-r^2/\sigma^2\right)}}{r}\int_0^R
\left[\exp\left(-\frac{r^{\prime 2}-2rr^\prime}{\sigma^2}\right)
-\exp\left(-\frac{r^{\prime
2}+2rr^\prime}{\sigma^2}\right)\right]r^\prime\,dr^\prime.
\end{equation}
Integrating over all the possible final positions corresponding to an overlap between the particles ($0\!\leq\! r^\prime\!\leq\! R$), gives
\begin{equation}
g(r)=\frac{\sigma}{\sqrt{\pi}}\frac{1}{2r}\left[\exp\left(-\frac{(r+R)^2}{\sigma^2}\right)-
\exp\left(-\frac{(r-R)^2}{\sigma^2}\right)\right]+
\frac{1}{2}\left[\rm{erf}\left(\frac{r+R}{\sigma}\right)+\rm{erf}\left(\frac{-r+R}{\sigma}\right)
\right],
\end{equation}
where
$$
\rm{erf}(x)=\frac{2}{\sqrt{\pi}}\int_0^x e^{t^2/2}dt.
$$

\begin{acknowledgments}
The authors are grateful to Sorin T\u{a}nase-Nicola for many valuable discussions. This work is part of the research program of the
``Stichting voor Fundamenteel Onderzoek der Materie (FOM)", which is financially supported by the "Nederlandse organisatie voor
Wetenschappelijk Onderzoek (NWO)''.
\end{acknowledgments}

\bibliography{BD_7}

\begin{thebibliography}{25}
\expandafter\ifx\csname natexlab\endcsname\relax\def\natexlab#1{#1}\fi
\expandafter\ifx\csname bibnamefont\endcsname\relax
  \def\bibnamefont#1{#1}\fi
\expandafter\ifx\csname bibfnamefont\endcsname\relax
  \def\bibfnamefont#1{#1}\fi
\expandafter\ifx\csname citenamefont\endcsname\relax
  \def\citenamefont#1{#1}\fi
\expandafter\ifx\csname url\endcsname\relax
  \def\url#1{\texttt{#1}}\fi
\expandafter\ifx\csname urlprefix\endcsname\relax\def\urlprefix{URL }\fi
\providecommand{\bibinfo}[2]{#2}
\providecommand{\eprint}[2][]{\url{#2}}

\bibitem[{\citenamefont{Ermak and McCammon}(1978)}]{Ermak78}
\bibinfo{author}{\bibfnamefont{D.~L.} \bibnamefont{Ermak}} \bibnamefont{and}
  \bibinfo{author}{\bibfnamefont{J.~A.} \bibnamefont{McCammon}},
  \bibinfo{journal}{J. Chem. Phys.} \textbf{\bibinfo{volume}{69}},
  \bibinfo{pages}{1352} (\bibinfo{year}{1978}).

\bibitem[{\citenamefont{Northrup and Erickson}(1992)}]{Northrup92}
\bibinfo{author}{\bibfnamefont{S.~H.} \bibnamefont{Northrup}} \bibnamefont{and}
  \bibinfo{author}{\bibfnamefont{H.~P.} \bibnamefont{Erickson}},
  \bibinfo{journal}{Proc. Natl. Acad. Sci. USA} \textbf{\bibinfo{volume}{89}},
  \bibinfo{pages}{3338} (\bibinfo{year}{1992}).

\bibitem[{\citenamefont{Wade}(1996)}]{Wade96}
\bibinfo{author}{\bibfnamefont{R.~C.} \bibnamefont{Wade}},
  \bibinfo{journal}{Biochem. Soc. Trans.} \textbf{\bibinfo{volume}{24}},
  \bibinfo{pages}{254} (\bibinfo{year}{1996}).

\bibitem[{\citenamefont{Huber and Kim}(1996)}]{Huber96}
\bibinfo{author}{\bibfnamefont{G.~A.} \bibnamefont{Huber}} \bibnamefont{and}
  \bibinfo{author}{\bibfnamefont{S.}~\bibnamefont{Kim}},
  \bibinfo{journal}{Biophys. J.} \textbf{\bibinfo{volume}{70}},
  \bibinfo{pages}{97} (\bibinfo{year}{1996}).

\bibitem[{\citenamefont{Gabdoulline and Wade}(1997)}]{Gabdoulline97}
\bibinfo{author}{\bibfnamefont{R.~R.} \bibnamefont{Gabdoulline}}
  \bibnamefont{and} \bibinfo{author}{\bibfnamefont{R.~C.} \bibnamefont{Wade}},
  \bibinfo{journal}{Biophys. J.} \textbf{\bibinfo{volume}{72}},
  \bibinfo{pages}{1917} (\bibinfo{year}{1997}).

\bibitem[{\citenamefont{Zou et~al.}(2000)\citenamefont{Zou, Skeel, and
  Subramaniam}}]{Zou00}
\bibinfo{author}{\bibfnamefont{G.}~\bibnamefont{Zou}},
  \bibinfo{author}{\bibfnamefont{R.~D.} \bibnamefont{Skeel}}, \bibnamefont{and}
  \bibinfo{author}{\bibfnamefont{S.}~\bibnamefont{Subramaniam}},
  \bibinfo{journal}{Biophys. J.} \textbf{\bibinfo{volume}{79}},
  \bibinfo{pages}{638} (\bibinfo{year}{2000}).

\bibitem[{\citenamefont{Gabdoulline and Wade}(2001)}]{Gabdoulline01}
\bibinfo{author}{\bibfnamefont{R.~R.} \bibnamefont{Gabdoulline}}
  \bibnamefont{and} \bibinfo{author}{\bibfnamefont{R.~C.} \bibnamefont{Wade}},
  \bibinfo{journal}{J. Mol. Biol.} \textbf{\bibinfo{volume}{306}},
  \bibinfo{pages}{1139} (\bibinfo{year}{2001}).

\bibitem[{\citenamefont{Gabdoulline and Wade}(2002)}]{Gabdoulline02}
\bibinfo{author}{\bibfnamefont{R.~R.} \bibnamefont{Gabdoulline}}
  \bibnamefont{and} \bibinfo{author}{\bibfnamefont{R.~C.} \bibnamefont{Wade}},
  \bibinfo{journal}{Curr. Opin. Struct. Biol.} \textbf{\bibinfo{volume}{12}},
  \bibinfo{pages}{204} (\bibinfo{year}{2002}).

\bibitem[{\citenamefont{Schaff et~al.}(1997)\citenamefont{Schaff, Fink,
  Slepchenko, Carson, and Loew}}]{VirtualCell}
\bibinfo{author}{\bibfnamefont{J.}~\bibnamefont{Schaff}},
  \bibinfo{author}{\bibfnamefont{C.~C.} \bibnamefont{Fink}},
  \bibinfo{author}{\bibfnamefont{B.}~\bibnamefont{Slepchenko}},
  \bibinfo{author}{\bibfnamefont{J.~H.} \bibnamefont{Carson}},
  \bibnamefont{and} \bibinfo{author}{\bibfnamefont{L.~M.} \bibnamefont{Loew}},
  \bibinfo{journal}{Biophys. J.} \textbf{\bibinfo{volume}{73}},
  \bibinfo{pages}{1135} (\bibinfo{year}{1997}).

\bibitem[{\citenamefont{Andrews and Bray}(2004)}]{Smoldyn}
\bibinfo{author}{\bibfnamefont{S.~S.} \bibnamefont{Andrews}} \bibnamefont{and}
  \bibinfo{author}{\bibfnamefont{D.}~\bibnamefont{Bray}},
  \bibinfo{journal}{Physical Biology} \textbf{\bibinfo{volume}{1}},
  \bibinfo{pages}{137} (\bibinfo{year}{2004}).

\bibitem[{\citenamefont{Stiles}(2000)}]{Mcell}
\bibinfo{author}{\bibfnamefont{J.~R.} \bibnamefont{Stiles}},
  \emph{\bibinfo{title}{Computational Neuroscience: Realistic Modeling for
  Experimentalists}} (\bibinfo{publisher}{CRC Press}, \bibinfo{address}{Boca
  Raton}, \bibinfo{year}{2000}).

\bibitem[{\citenamefont{Gillespie}(1977)}]{Gillespie77}
\bibinfo{author}{\bibfnamefont{D.~T.} \bibnamefont{Gillespie}},
  \bibinfo{journal}{J. Phys. Chem.} \textbf{\bibinfo{volume}{81}},
  \bibinfo{pages}{2340 } (\bibinfo{year}{1977}).

\bibitem[{\citenamefont{Hattne et~al.}(2005)\citenamefont{Hattne, Fange, and
  Elf}}]{MesoRD}
\bibinfo{author}{\bibfnamefont{J.}~\bibnamefont{Hattne}},
  \bibinfo{author}{\bibfnamefont{D.}~\bibnamefont{Fange}}, \bibnamefont{and}
  \bibinfo{author}{\bibfnamefont{J.}~\bibnamefont{Elf}},
  \bibinfo{journal}{Bioinformatics} \textbf{\bibinfo{volume}{21}},
  \bibinfo{pages}{2923} (\bibinfo{year}{2005}).

\bibitem[{\citenamefont{Ander et~al.}(2004)\citenamefont{Ander, Beltrao,
  Ventura, Ferkinghoff-Berg, Foglierini, Kaplan, Lemerle, Toma's-Oliveira, and
  Serrano}}]{SmartCell}
\bibinfo{author}{\bibfnamefont{M.}~\bibnamefont{Ander}},
  \bibinfo{author}{\bibfnamefont{P.}~\bibnamefont{Beltrao}},
  \bibinfo{author}{\bibfnamefont{B.~D.} \bibnamefont{Ventura}},
  \bibinfo{author}{\bibfnamefont{J.}~\bibnamefont{Ferkinghoff-Berg}},
  \bibinfo{author}{\bibfnamefont{M.}~\bibnamefont{Foglierini}},
  \bibinfo{author}{\bibfnamefont{A.}~\bibnamefont{Kaplan}},
  \bibinfo{author}{\bibfnamefont{C.}~\bibnamefont{Lemerle}},
  \bibinfo{author}{\bibfnamefont{I.}~\bibnamefont{Toma's-Oliveira}},
  \bibnamefont{and} \bibinfo{author}{\bibfnamefont{L.}~\bibnamefont{Serrano}},
  \bibinfo{journal}{Syst. Biol.} \textbf{\bibinfo{volume}{1}},
  \bibinfo{pages}{129} (\bibinfo{year}{2004}).

\bibitem[{\citenamefont{Lemerle et~al.}(2005)\citenamefont{Lemerle, Ventura,
  and Serrano}}]{Lemerle05}
\bibinfo{author}{\bibfnamefont{C.}~\bibnamefont{Lemerle}},
  \bibinfo{author}{\bibfnamefont{B.~D.} \bibnamefont{Ventura}},
  \bibnamefont{and} \bibinfo{author}{\bibfnamefont{L.}~\bibnamefont{Serrano}},
  \bibinfo{journal}{FEBS Letters} \textbf{\bibinfo{volume}{579}},
  \bibinfo{pages}{1789} (\bibinfo{year}{2005}).

\bibitem[{\citenamefont{Rodriguez et~al.}(2006)\citenamefont{Rodriguez,
  Kaandorp, Dobrzynski, and Blom}}]{Rodriguez06}
\bibinfo{author}{\bibfnamefont{J.~V.} \bibnamefont{Rodriguez}},
  \bibinfo{author}{\bibfnamefont{J.~A.} \bibnamefont{Kaandorp}},
  \bibinfo{author}{\bibfnamefont{M.}~\bibnamefont{Dobrzynski}},
  \bibnamefont{and} \bibinfo{author}{\bibfnamefont{J.~G.} \bibnamefont{Blom}},
  \bibinfo{journal}{Bioinformatics} \textbf{\bibinfo{volume}{22}},
  \bibinfo{pages}{1895} (\bibinfo{year}{2006}).

\bibitem[{\citenamefont{Agmon and Szabo}(1990)}]{Agmon90}
\bibinfo{author}{\bibfnamefont{N.}~\bibnamefont{Agmon}} \bibnamefont{and}
  \bibinfo{author}{\bibfnamefont{A.}~\bibnamefont{Szabo}}, \bibinfo{journal}{J.
  Chem. Phys.} \textbf{\bibinfo{volume}{92}}, \bibinfo{pages}{5270}
  (\bibinfo{year}{1990}).

\bibitem[{\citenamefont{van Zon et~al.}(2006)\citenamefont{van Zon, Morelli,
  Tanase-Nicola, and ten Wolde}}]{VanZon06}
\bibinfo{author}{\bibfnamefont{J.~S.} \bibnamefont{van Zon}},
  \bibinfo{author}{\bibfnamefont{M.~J.} \bibnamefont{Morelli}},
  \bibinfo{author}{\bibfnamefont{S.}~\bibnamefont{Tanase-Nicola}},
  \bibnamefont{and} \bibinfo{author}{\bibfnamefont{P.~R.} \bibnamefont{ten
  Wolde}}, \bibinfo{journal}{Biophys. J.} \textbf{\bibinfo{volume}{91}},
  \bibinfo{pages}{4350} (\bibinfo{year}{2006}).

\bibitem[{\citenamefont{Frenkel and Smit}(2002)}]{Frenkelbook}
\bibinfo{author}{\bibfnamefont{D.}~\bibnamefont{Frenkel}} \bibnamefont{and}
  \bibinfo{author}{\bibfnamefont{B.}~\bibnamefont{Smit}},
  \emph{\bibinfo{title}{Understanding Molecular Simulations: From Algorithms to
  Applications, 2nd ed.}} (\bibinfo{publisher}{Academic},
  \bibinfo{address}{Boston}, \bibinfo{year}{2002}).

\bibitem[{\citenamefont{Press et~al.}(1992)\citenamefont{Press, Teukolsky,
  Vetterling, and Flannery}}]{NumRec}
\bibinfo{author}{\bibfnamefont{W.~H.} \bibnamefont{Press}},
  \bibinfo{author}{\bibfnamefont{S.}~\bibnamefont{Teukolsky}},
  \bibinfo{author}{\bibfnamefont{W.~T.} \bibnamefont{Vetterling}},
  \bibnamefont{and} \bibinfo{author}{\bibfnamefont{B.~P.}
  \bibnamefont{Flannery}}, \emph{\bibinfo{title}{Numerical Recipes in C, 2nd
  ed.}} (\bibinfo{publisher}{Cambridge University Press},
  \bibinfo{address}{Oakleigh}, \bibinfo{year}{1992}).

\bibitem[{\citenamefont{van Zon and ten Wolde}(2005)}]{VanZon05}
\bibinfo{author}{\bibfnamefont{J.~S.} \bibnamefont{van Zon}} \bibnamefont{and}
  \bibinfo{author}{\bibfnamefont{P.~R.} \bibnamefont{ten Wolde}},
  \bibinfo{journal}{Phys. Rev. Lett.} \textbf{\bibinfo{volume}{94}},
  \bibinfo{pages}{018104} (\bibinfo{year}{2005}).

\bibitem[{\citenamefont{Zon and ten Wolde}(2005)}]{VanZon05_2}
\bibinfo{author}{\bibfnamefont{J.~S.~V.} \bibnamefont{Zon}} \bibnamefont{and}
  \bibinfo{author}{\bibfnamefont{P.~R.} \bibnamefont{ten Wolde}},
  \bibinfo{journal}{J. Chem. Phys.} \textbf{\bibinfo{volume}{123}},
  \bibinfo{pages}{234910} (\bibinfo{year}{2005}).

\bibitem[{\citenamefont{Kim and Shin}(1998)}]{Kim98}
\bibinfo{author}{\bibfnamefont{H.}~\bibnamefont{Kim}} \bibnamefont{and}
  \bibinfo{author}{\bibfnamefont{K.~J.} \bibnamefont{Shin}},
  \bibinfo{journal}{Phys. Rev. Lett.} \textbf{\bibinfo{volume}{82}},
  \bibinfo{pages}{1578 } (\bibinfo{year}{1998}).

\bibitem[{\citenamefont{Goldbeter and D.~E.~Koshland}(1981)}]{Goldbeter81}
\bibinfo{author}{\bibfnamefont{A.}~\bibnamefont{Goldbeter}} \bibnamefont{and}
  \bibinfo{author}{\bibfnamefont{J.}~\bibnamefont{D.~E.~Koshland}},
  \bibinfo{journal}{Proc. Natl. Acad. Sci. USA} \textbf{\bibinfo{volume}{78}},
  \bibinfo{pages}{6840} (\bibinfo{year}{1981}).

\bibitem[{\citenamefont{Berg et~al.}(2000)\citenamefont{Berg, Paulsson, and
  Ehrenberg}}]{Berg00}
\bibinfo{author}{\bibfnamefont{O.~G.} \bibnamefont{Berg}},
  \bibinfo{author}{\bibfnamefont{J.}~\bibnamefont{Paulsson}}, \bibnamefont{and}
  \bibinfo{author}{\bibfnamefont{M.}~\bibnamefont{Ehrenberg}},
  \bibinfo{journal}{Biophys. J.} \textbf{\bibinfo{volume}{79}},
  \bibinfo{pages}{1228} (\bibinfo{year}{2000}).

\end{thebibliography}

\newpage 

\begin{center} 
\textbf{LIST OF FIGURES} 
\end{center}

\begin{enumerate}

\item Radial probability distribution for an {\em irreversible}
reaction. The four curves refer to different $t_{\rm{sim}}$ and were
obtained with time steps $\Delta t\!=\!10^{-4}t_{\rm{sim}}$, except for $t_{\rm sim}\!=\!0.1\tau$ where we used $\Delta
t\!=\!10^{-5}t_{\rm{sim}}\!=\!10^{-6}\tau$
($\tau\!=\!R^2/D$, $R\!=\!R_A\!+\!R_B$). Particles were initially positioned at contact: $r_0\!=\!R$. The intrinsic association constant is
$k_{\rm a}\!=\!1000 R^3/\tau$. 
The numerical results (symbols) are in excellent agreement with the
analytical curves (solid lines) \cite{Kim98}.
In the Inset, we plot the probability distribution for $t_{\rm sim}\!=\!10^{-1}\tau$ for several time steps. 
For large $\Delta t$ the BD algorithm deviates from the analytical line and 
underestimates the survival probability. Error bars are smaller than symbol sizes. 
\label{fig:pirr}

\item Radial probability distribution for a {\em reversible}
reaction. The four curves refer to different $t_{\rm{sim}}$ and were
obtained with time steps $t_{\rm{step}}\!=\!10^{-4}t_{\rm{sim}}$,
$r_0\!=\!R$, except for $t_{\rm sim}\!=\!0.1\tau$ where we used $\Delta
t\!=\!10^{-5}t_{\rm{sim}}\!=\!10^{-6}\tau$ ($\tau\!=\!R^2/D$, $R\!=\!R_A+R_B$). 
Particles were initially positioned at contact ($r_0\!=\!R$), and the
association rate constant is 
$k_{\rm a}\!=\!1000 R^3/\tau$ ($\tau\!=\!R^2/D$, $R\!=\!R_A+R_B$),
while the dissociation rate constant is set to $k_{\rm d}\!=\!1\tau^{-1}$. 
The numerical results (circles) agree with the analytical curves (solid lines).
In the Inset, the probability distribution for $t_{\rm sim}\!=\!10^{-1}\tau$ is plotted for several values of  $\Delta t$: for large 
values of the time step, the BD algorithm deviates from the analytical line and underestimates the survival probability. Error bars are
smaller than symbol sizes. 
\label{fig:prev}

\item Probability of having an $A$ particle bound to a $B$
  particle, as a function of the number of $B$ particles.  The time
  step is set to $\Delta t\!=\!10^{-4}\tau$ ($\tau\!=\!R^2/D$,
  $R\!=\!R_A+R_B$), the intrinsic association constant to $k_{\rm
    on}\!=\!71R^3/\tau$ so that $P_{\rm acc,\,f}\!=\!0.1$.  $k_{\rm
    b}$ is chosen so that $K_{\rm eq}\!=\!k_{\rm f}/k_{\rm b}\!=\!V$
  ($V\!=\!8000R^3$) and therefore $p_{\rm bound}(N_B\!=\!1)\!=\!0.5$.
  The numerical data obtained with BD are in agreement with the
  mean-field values. The error bars of the numerical results are
  smaller than the size of the circles.  In Inset A, the simulations
  are performed positioning dissociated particles at contact. This
  move violates detailed balance and yield an incorrect $p_{\rm
    bound}$ for low number of particles.  In Inset B, $p_{\rm bound}$,
  for $N_B\!=\!1$ is plotted against the time step used in the
  simulations. To keep $P_{\rm acc,\,f}\!=\!0.1$, we varied $k_{\rm
    on}$ from $2242R^3/\tau$ ($\Delta t\!=\!10^{-7}\tau$) to
  $0.00026R^3/\tau$ ($\Delta t\!=\!10^{-1}\tau$). As expected for an
  equilibrium quantity, $p_{\rm bound}$ does not depend on the chosen
  time step. 
 \label{fig:p_bound_NB}
 
\item Distribution of association times, for the reaction
$A+B\leftrightarrow C$, obtained with the Brownian Dynamics
algorithm (solid line) and with a Stochastic Simulation Algorithm
(dashed line) neglecting spatial effects. The data are obtained for
$V\!=\!64000R^3,D\!=\!R^2/\tau,\Delta t\!=\!10^{-4}\tau,k_{\rm a}\!=\!100R^3/\tau,k_{\rm d}\!=\!1000\tau^{-1}$ ($\tau\!=\!R^2/D$, $R\!=\!R_A+R_B$).
Spatial simulations account for immediate rebindings after a
dissociation event, and show a higher probability for short
association times. The two curves decay exponentially to zero with
the same rate  $k_{\rm f}^{-1}=k_{\rm a}^{-1}\!+\!k_{\rm D}^{-1}$. 
\label{fig:treact}

\item Snapshot of the push-pull system. The pink and the green
sphere represent, respectively, the kinase and the phosphatase
molecule, held fixed along the main axis of the box. The blue and
red spheres represent $S$ and $S_p$ molecules, respectively. The
system is represented for $N_{S\rm tot}\!=\!50$ and a box of
$20R\times10R\times10R$. 
\label{fig:snapshot}

\item Fraction of converted molecules as a function of ratio of the
  convertion rates $k_1/k_2$. The data are shown for increasing values
  of $K_{\rm M}/[S_{\rm tot}]$ from panel A to D. Panel A corresponds to full
  saturation of enzymes, whereas in panel D the system is in the
  first-order regime.  The continuous lines are obtained by solving
  the Macroscopical Rate Equation, whereas diamonds correspond to the
  numerical solutions of the master equation (obtained with the
  conventional SSA) and circles to the output of our BD
  simulations. SSA data (error bars are smaller than the sizes of the
  symbols) show a mild deviation only when the system displays an
  ultrasensitive behavior, as in panel A. Brownian
  Dynamics simulations deviate notably from the macroscopic curve
  when $K_{\rm M}/[S_{\rm tot}]\!<\!1$.  Methods accounting for the stochastic
  behavior of the system show thus a reduction in sensitivity for low
  values of $K_{\rm M}\!$. In the simulations, we vary $k_1$ and $k_{\rm off2}$ keeping their
  sum constant, so that $k_1/k_2$ is varied while $K_{\rm M}$ does not change. For all
  panels, $D=10^{-3}R^2/(\Delta t),N_{S_{\rm tot}}=80$. 
  Panel A: $k_1+k_{\rm off1}=k_2+k_{\rm off2}=5\cdot 10^{-6}\Delta t^{-1}$, $k_{\rm a}=0.006R^3/(\Delta t)$. 
  Panel B: $k_1+k_{\rm off1}=k_2+k_{\rm off2}=5\cdot 10^{-6}\Delta t^{-1}$, $k_{\rm a}=0.001R^3/(\Delta t)$,  
  Panel C: $k_1+k_{\rm off1}=k_2+k_{\rm off2}=5\cdot 10^{-5}\Delta t^{-1}$, $k_{\rm a}=0.003R^3/(\Delta t)$,  
  Panel D: $k_1+k_{\rm off1}=k_2+k_{\rm off2}=2\cdot 10^{-4}\Delta t^{-1}$, $k_{\rm a}=0.0015R^3/(\Delta t)$.  
  \label{fig:sensitivity}
  
\item The spatial density profiles for $S$, $S_p$ ($k_1/k_2\!=\!1$) show clear symmetric peaks around the locations of the two enzymes:
respectively, the $S$ density is peaked around the kinase enzyme, at $x\!=\!0.25x_{\rm box}$, and the $S_p$ density around the phosphatase at
$x\!=\!-0.25x_{\rm box}$. These peaks are more pronounced when the system is in the ultrasensitive regime (thinner lines). 
Moreover, for high values of the Michaelis-Menten constants (thicker lines), the spatial density of the particles show a gradients, higher
close to the production sites of the molecules. For low values of $K_{\rm M}$ the gradients are not appreciable anymore: 
the particles can diffuse in the box and reach an homogeneous distribution, as production events happen on slow time scales.
Simulation parameters:  
$K_{\rm M}/[S_{\rm tot}]=4.7,D=10^{-4}R^2/(\Delta t),N_{S_{\rm tot}}=80,k_1=k_2=k_{\rm off1}=k_{\rm off2}=10^{-4} \Delta t^{-1},
k_{\rm a}=0.007 R^3/(\Delta t)$
$K_{\rm M}/[S_{\rm tot}]=0.22,D=10^{-3}R^2/(\Delta t),N_{S_{\rm tot}}=80,k_1=k_2=k_{\rm off1}=k_{\rm off2}=2.5 \cdot 10^{-5} \Delta t^{-1},
k_{\rm a}=0.01 R^3/(\Delta t)$
\label{fig:grad}

\end{enumerate}

\newpage
\vspace*{2cm} 
\begin{figure}[h] 
\includegraphics[width=0.8\textwidth]{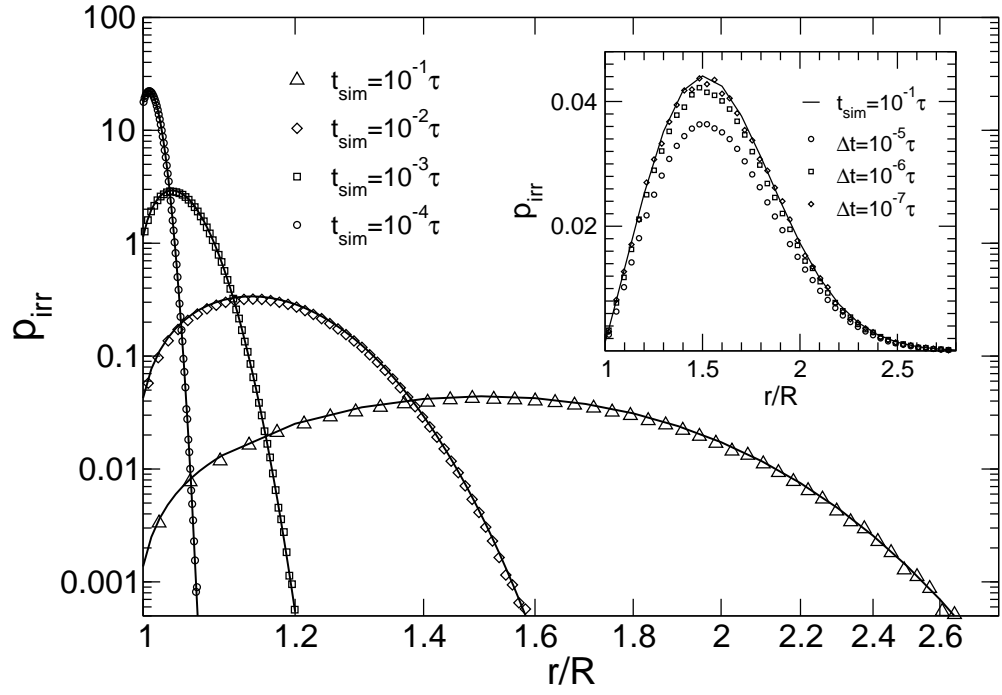}
\caption{Morelli and ten Wolde}
\end{figure}

\newpage
\vspace*{2cm} 
\begin{figure}
\includegraphics[width=0.8\textwidth]{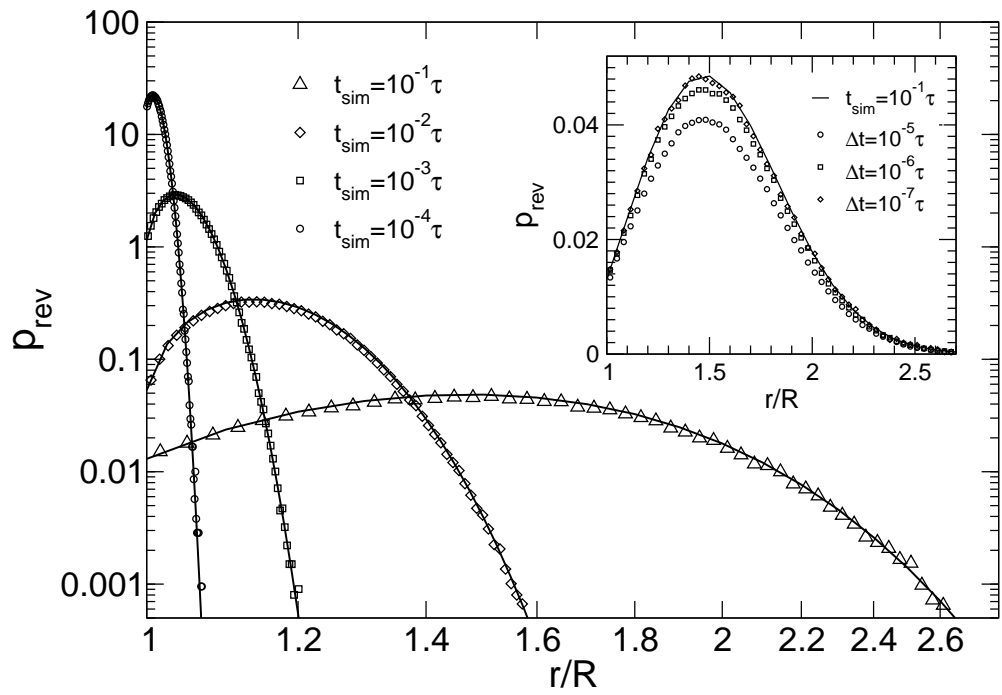}
\caption{Morelli and ten Wolde}
\end{figure}

\newpage
\vspace*{2cm} 
\begin{figure}
\includegraphics[width=0.8\textwidth]{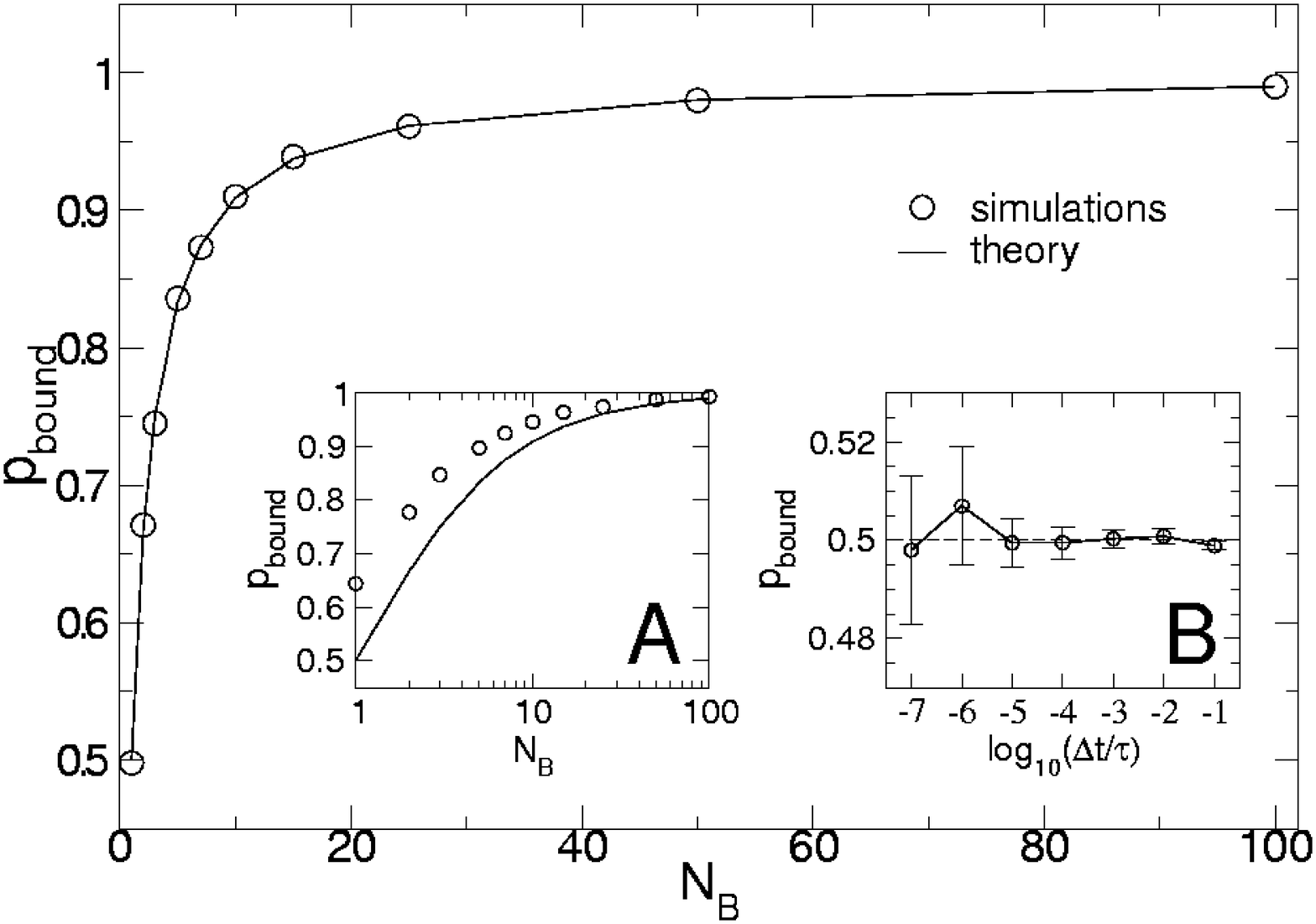}
\caption{Morelli and ten Wolde}
\end{figure}

\newpage
\vspace*{2cm} 
\begin{figure}
\includegraphics[width=0.8\textwidth]{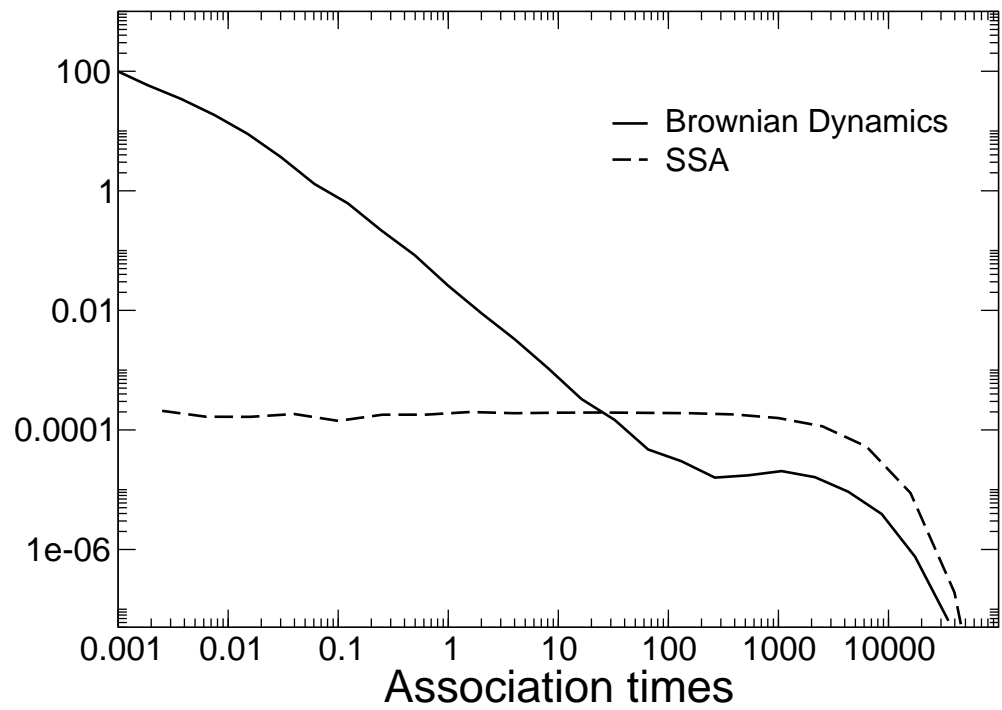}
\caption{Morelli and ten Wolde}
\end{figure}

\newpage
\vspace*{2cm} 
\begin{figure}
\includegraphics[width=0.8\textwidth]{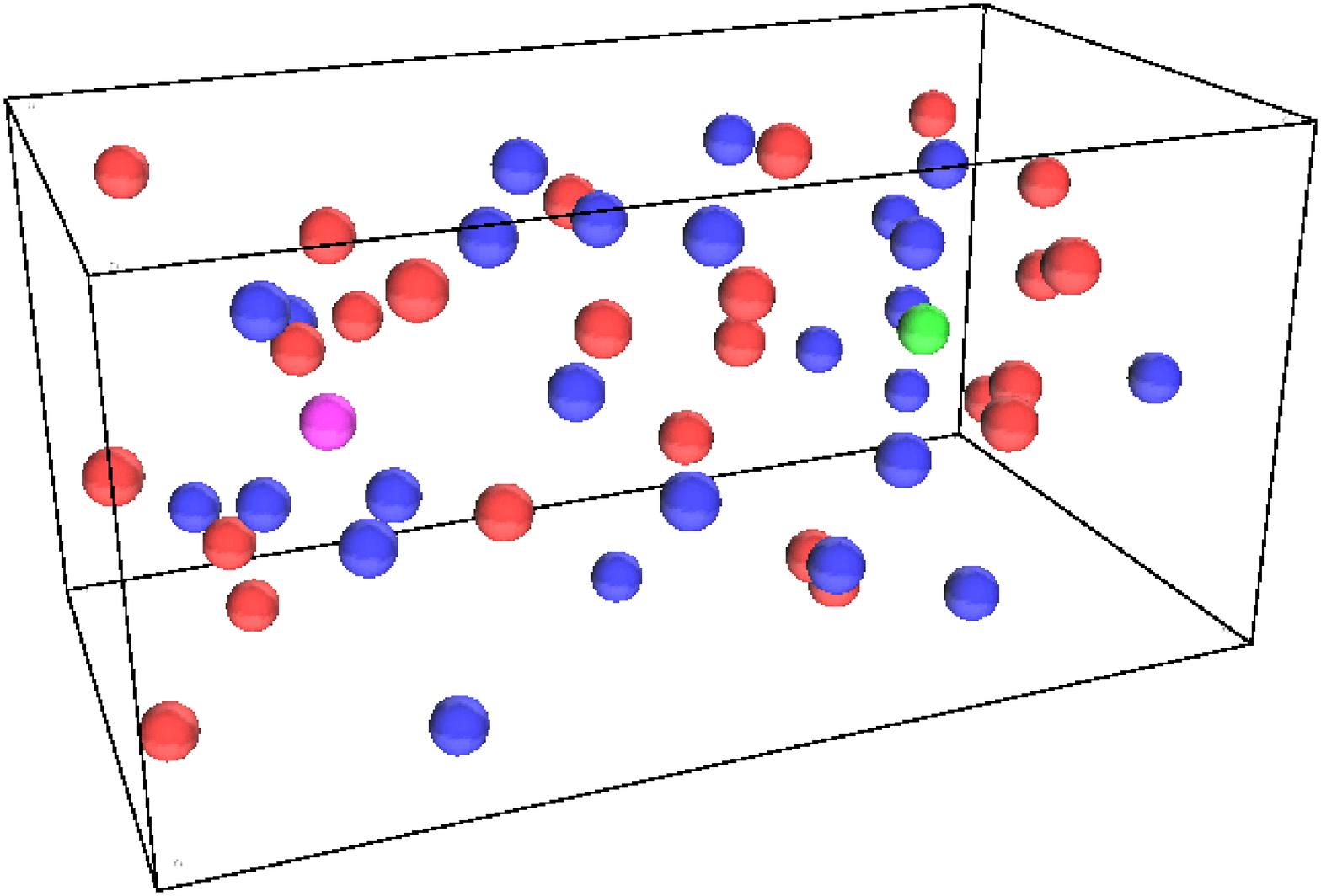}
\caption{Morelli and ten Wolde}
\end{figure}

\newpage
\vspace*{2cm} 
\begin{figure}
\includegraphics[width=0.8\textwidth]{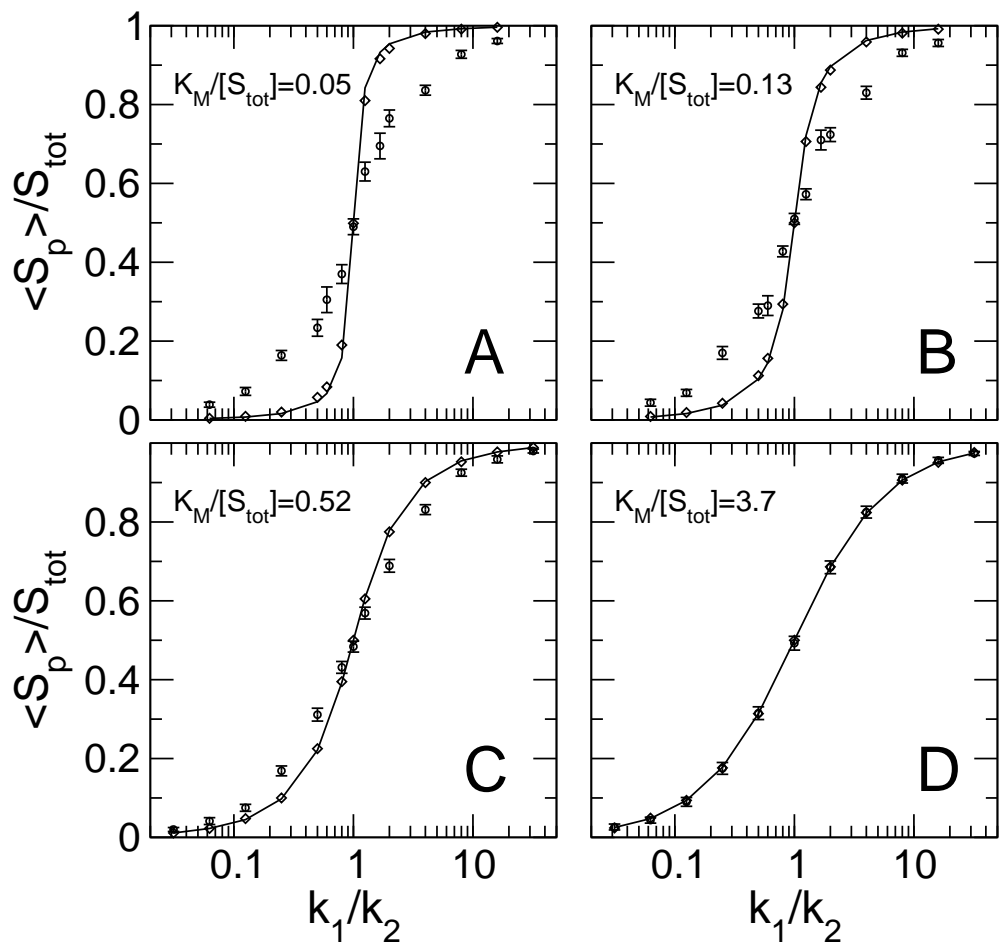}
\caption{Morelli and ten Wolde}
\end{figure}

\newpage
\vspace*{2cm} 
\begin{figure}
\includegraphics[width=0.8\textwidth]{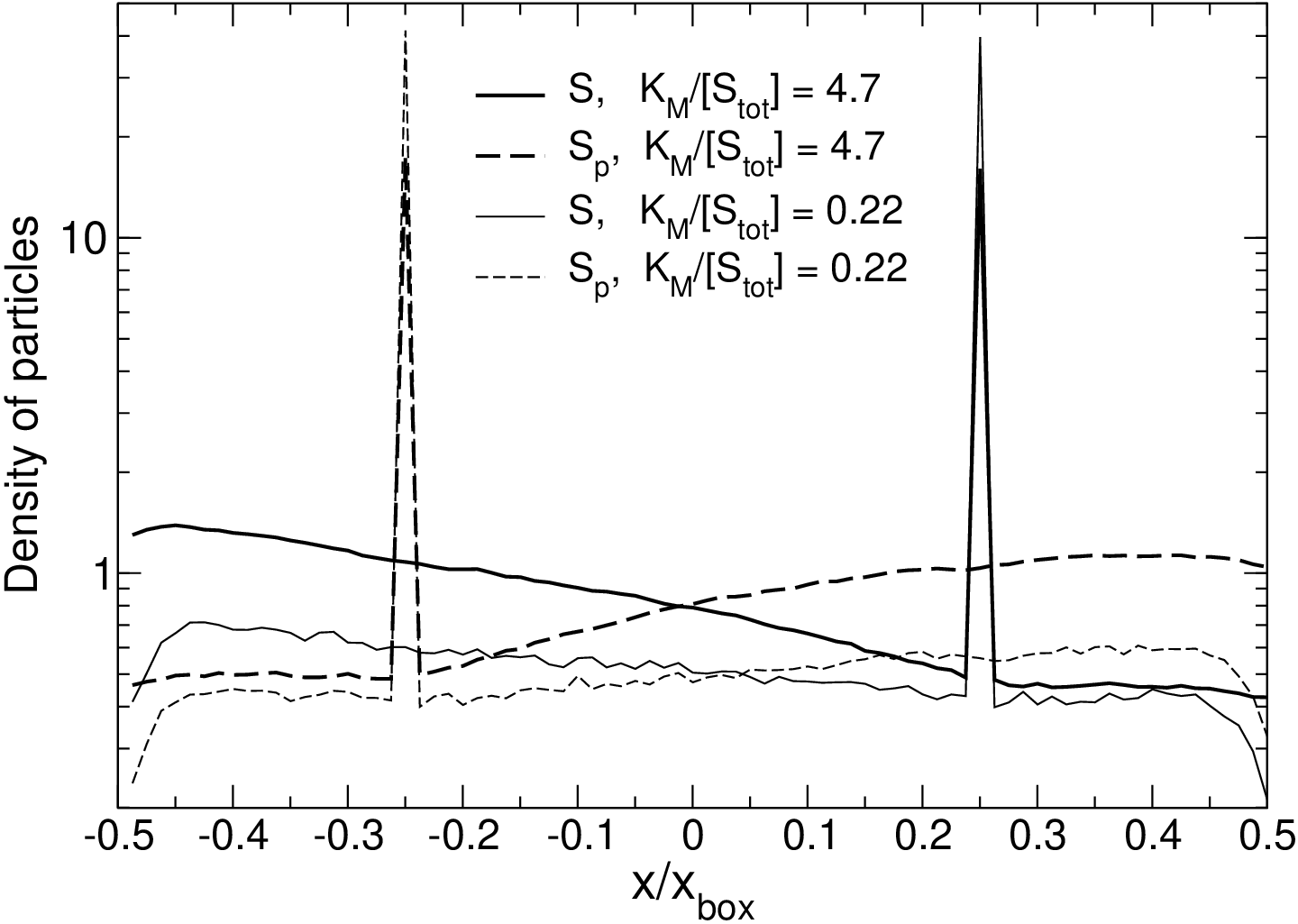}
\caption{Morelli and ten Wolde}
\end{figure}

\end{document}